\documentclass[preprint,3p]{elsarticle}
\usepackage{longtable}
\usepackage{listings}
\usepackage{hyperref}
\usepackage{booktabs}
\usepackage{subcaption}
\usepackage{color}
\usepackage{amssymb}
\usepackage{amsmath}
\usepackage{appendix}
\usepackage{xcolor}
\usepackage{multicol}
\usepackage{float}
\usepackage[T1]{fontenc}
\usepackage{array}
\usepackage{multirow}

\usepackage{algorithmic}
\usepackage{graphicx}
\usepackage{textcomp}
\usepackage{stfloats}
\usepackage{placeins}
\usepackage[ruled,vlined]{algorithm2e}

\usepackage[normalem]{ulem}

\newtheorem{Ex}{Example}

\newtheorem{proofhead}{Proof}

\biboptions{sort&compress}
\journal{Statistical Papers}

\begin{document}

\begin{frontmatter}



\title{
A systematic framework for the identification and statistical quantification of impulsivity in condition monitoring signals} 
\author[inst1]{Aleksandra Grzesiek}
\author[inst1]{Justyna Witulska\corref{cor1}}
\author[inst2]{Daniel Kuzio}
\author[inst2]{Radosław Zimroz}
\author[inst3]{Tomasz Barszcz}
\author[inst1]{Agnieszka Wyłomańska}

\cortext[cor1]{Corresponding author. Email: justyna.witulska@pwr.edu.pl}

\affiliation[inst1]{organization={Faculty of Pure and Applied Mathematics, Hugo Steinhaus Center, Wroclaw University of Science and Technology},
            addressline={Hoene-Wronskiego 13c},
            city={Wroclaw},
            postcode={50-376},
                 country={Poland}}
                                  \affiliation[inst2]{organization={Faculty of Geoengineering, Mining and Geology, Wroclaw University of Science and Technology},
            addressline={Na Grobli 15},
            city={Wroclaw},
            postcode={50-421},
            country={Poland}}
                              \affiliation[inst3]{organization={Faculty of Mechanical Engineering and Robotics, Department of Robotics and Mechatronics, AGH University of Krakow},
            addressline={al. A. Mickiewicza 30},
            city={Krakow},
            postcode={30-059},
            country={Poland}}
                    
         \begin{abstract}
This article proposes a comprehensive framework for the identification and statistical quantification of impulsive behavior in signals, with a primary focus on condition monitoring. We concentrate on evaluating impulsivity, where such behavior results from normal operation or additional disturbances. Such an evaluation is crucial in the context of local damage detection, as the presence of impulsive disturbances significantly complicates the machine condition monitoring process. To address problem of impulsivity assessment
we introduce a two-stage methodology to  make processing workflow effective. First, we propose an objective selection criterion for "best-performing" impulsivity measures based on the Mann-Whitney statistic, allowing for the systematic comparison of various classical and advanced metrics across diverse signal scenarios. Second, we establish a formal procedure for assessing statistical significance using bootstrap-driven resampling and define a magnitude index to quantify the intensity of detected impulsivity. The framework is validated through extensive Monte Carlo simulations for three reference signal scenarios and applied to real-world vibration data from an industrial compressor. By systematizing existing measures and providing a statistically grounded pipeline, this research extends prior works, offering a scalable tool for distinguishing between diagnostically useful signals and those corrupted by anomalous interference.
\end{abstract}


\end{frontmatter}

\section{Introduction and motivation}\label{intro}
This article addresses the problem of assessing and parameterizing impulsive behavior in signals. While our primary motivation stems from the field of condition monitoring and the evaluation of impulsivity in diagnostic signals (e.g., vibration data), the underlying issue has broader implications. The methodology proposed in this article is applicable across any domain where signals or data exhibit impulsive characteristics. In the specific context of condition monitoring, we focus on evaluating impulsivity where 
such behavior results from normal operation as well as additional disturbances, which often complicates the monitoring process.

In signal processing literature and practical applications, the challenge of impulsive (non-Gaussian) data is increasingly recognized, see Section \ref{inne} for more details. Consequently, robust methods are being proposed to replace classical approaches designed for Gaussian signals; see Section \ref{inne1} for the condition monitoring area. It is evident that analyzing impulsive signals requires advanced techniques 
that are resilient to impulsive and anomalous (outliers) behaviors. On the other hand, robust methods are significantly more time-consuming. Therefore, for signals exhibiting "normal", (i.e non-impulsive) behavior, advanced analysis is often unnecessary. This creates a clear need for a preliminary assessment stage to detect the impulsivity of the given signal (binary classification: impulsive vs. non-impulsive) and to  quantify the level of impulsivity to select the appropriate analytical tools, i.e., utilizing fast, classical methods for non-impulsive signals and more effective, albeit computationally intensive, robust methods for  impulsive ones. Furthermore, quantifying impulsivity  is vital for monitoring systems to identify signals that are excessively corrupted by impulsive noise to provide reliable diagnostic information.

The literature offers numerous classical metrics and advanced statistics (e.g., test statistics for Gaussian distribution testing) that serve as measures of impulsivity. 
These are briefly discussed in Section \ref{inne3}, see also  Section \ref{sec:measures} for more detailed discussion. However, a critical question remains: which measure is the most effective and reliable for a given reference scenario?  Since different metrics may return values in different ranges, a simple comparative analysis is often insufficient for selecting the optimal measure. Thus, there is a need for a universal methodology for  the  selection of the most efficient metric. Once a signal is identified as impulsive using the "optimal" measure, it is crucial to verify its statistical significance and level of impulsiveness. This quantification is key to classifying signals as diagnostically useful and deciding whether advanced robust methods are required.

In this study, we distinguish two  research problems:  (a) the selection of appropriate metrics for impulsiveness assessment, i.e.,  identifying measures that most accurately determine whether a signal contains anomalous impulsive interference; (b) the assessment of statistical significance, i.e., determining whether the detected impulsivity is statistically significant and defining its magnitude. Based on these two research problems, we designed a general framework consisting of two stages.

In the first stage of the proposed framework, namely the selection of the best-performing impulsivity measures, we propose the methodology based on the Mann-Whitney statistic \cite{mann1947test}, which provides a nonparametric measure of stochastic
distinction between two random variables. In our case, we adopt it for the discussed impulsivity measures calculated for reference and impulsive signals. This formulation allows for the objective evaluation of candidate impulsivity measures under diverse signal characteristics and for the identification of measures exhibiting high sensitivity to impulsive deviations in the assumed scenario. 

Then, statistical significance (the second stage of the proposed framework) is assessed using bootstrap resampling of the reference signal, from which one-sided confidence intervals are derived. A magnitude index, defined as the excess over the upper confidence bound, is introduced to quantify the strength of impulsivity.  The proposed framework ensures a consistent assessment and quantification of impulsive disturbances that can be useful in various contexts.

In the  simulation study presented, reference signal scenarios are defined to represent healthy machine at normal operating conditions, including stochastic Gaussian and autoregressive (AR) processes, as well as extended models incorporating cyclic components, similar to those observed in the vibration signals of the compressor. Reference signals are constructed by superimposing controlled impulsive disturbances, enabling systematic Monte Carlo experimentation across varying impulse amplitudes and varying numbers of impulses in the signal. 
The adopted scenarios are intended to describe the representative operating conditions observed in practice and to provide a controlled environment for systematic evaluation. It should be emphasized that the proposed assessment of impulsivity measures is not restricted to these specific signal models. The overall procedure is general and can be applied to alternative reference signals, provided that an appropriate baseline model and parameterization are defined. Consequently, the framework can be adaptable to different application domains and signal characteristics.

Finally, the proposed methodology is applied to real signals acquired from a compressor. The machine is a high-power reciprocating compressor used in the oil and gas industry. It consists of four pistons that progressively compress natural gas before it is transferred to the transportation pipeline. Piston operation generates complex vibration patterns with an impulsive character, even under normal operating conditions, while the signals are additionally influenced by varying operational regimes. Due to the harsh industrial environment, the measured signals often contain additional impulsive components. Some of these are caused by external events, whereas others result from factors such as cabling errors. Therefore, this case study is particularly well suited for investigating the impulsive characteristics of vibration signals. While the impulsive nature of the analyzed real signals is visually apparent, our methodology provides formal and objective confirmation.  The presented analysis of compressor vibration signals demonstrates the usefulness of the introduced framework in a real-world environment.

This article extends the research discussed in Skowronek et al. \cite{Skowronek2023}, where a simple statistic based on the cumulative sample fourth moment was proposed to assess signal impulsivity (here in the context of finite or infinite model variance). In \cite{Skowronek2023}, the aforementioned statistic was analyzed solely based on its chaotic behavior, and impulsivity (or the lack thereof) was identified via visual inspection. We also refer the readers to \cite{skowronek2}, where an attempt was made to quantify the chaotic behavior of empirical cumulative even moment statistics, also within the field of condition monitoring.  See also \cite{Hou2021199}, where various (classical) impulsivity measures are discussed in the context of fault detection and assessed using various evaluation metrics. The present study goes beyond the aforementioned results and represents their natural continuation.

The main novelty of this research lies in the development of a statistically grounded, two-stage selection and validation pipeline that eliminates the subjectivity typically inherent in choosing impulsivity measures for local damage detection. Moving beyond traditional heuristic selection, this framework introduces an objective criterion based on the Mann-Whitney statistic to identify the most robust metrics for specific signal environments. By integrating bootstrap-driven resampling to establish formal significance thresholds and a dedicated magnitude index, the methodology provides a rigorous means of distinguishing between "normal" behavior and impulsive anomalies. Additionally, this work systematizes existing measures by classifying them according to the statistical properties crucial for the characterization of impulsive background components. This also represents an additional novel contribution.

This research makes several significant contributions to the field of condition monitoring. First, it juxtaposes and systematizes a wide array of impulsivity measures (some of which were not previously utilized in condition monitoring), a critical step for both background noise characterization and impulsive signal assessment within fault detection procedures, particularly where faults manifest as impulses against near-Gaussian background noise. Second, it proposes a generalized methodology for selecting optimal measures for specific diagnostic scenarios, which is also directly applicable to fault detection. Third, it establishes a framework to evaluate the statistical significance and magnitude of identified impulsivity. In addition to the above mentioned novel contributions in the condition monitoring area, the introduction of heat maps of the discussed impulsivity measures and their objective evaluation reveals new knowledge about how such metrics behave under different values of amplitudes and different numbers of impulses in a given signal. While a compressor is utilized as a validation case study, the evaluation of impulsive background noise is relevant to a broad class of industrial processes (such as drilling, sieving, cutting, and grinding). Furthermore, condition monitoring in harsh environments often encounters external impulsive disturbances, such as electromagnetic interference or artifacts from wireless data transmission. Consequently, this work addresses a pervasive and challenging problem in the field of condition monitoring. 

The rest of the paper is organized as follows. In Section \ref{sota}, we present a brief state of the art, where we discuss different aspects of the problem. Specifically, we examine the issue of non-Gaussian (impulsive) data in various areas of interest, non-Gaussian-based techniques used in condition monitoring, as well as non-Gaussian-based models for industrial data. Finally, we indicate  impulsivity measures known across different fields. Next, in Section \ref{sec:measures}, we systematize the measures of impulsivity, discussing their main properties within the five proposed classes. 
In Section \ref{sec:methodology}, we describe the proposed framework, which consists of two steps mentioned earlier, namely the selection criteria and statistical evaluation of the detected impulsivity. In Section  \ref{sec:mc_sim}, the extensive simulation study for three adopted scenarios is discussed, which are simplified models of the vibration signals from healthy machines with possible impulses. Finally, in Section \ref{sec:real}, we present the analyses for real signals from the compressor. The last section summarizes the article. In the Appendices, we present the definitions of the discussed impulsivity measures, discuss the bootstrap-based methods used at the second stage of the proposed framework, and include additional results.
\section{State of the art}\label{sota}
The literature review presented below has been divided into several subsections that are closely related to the problem of impulsive signals and their processing.  
\subsection{Problem of non-Gaussian signals in different areas of interest}\label{inne}

While assessing impulsivity in condition monitoring is a relatively new development, addressing non-Gaussian interference is a long-standing challenge across diverse fields. For instance, Middleton \cite{Middleton1977106} explored the non-Gaussian nature of electromagnetic interference as early as the late 1970s. This challenge is also deeply rooted in fields such as general signal processing \cite{NAPOLITANO2025109980}, telecommunications \cite{Shongwe2015189}, and Internet of Things (IoT) data transmission \cite{Wang2023}. More recently, Zhou et al. \cite{Zhou2026} addressed the difficulties impulsive noise poses for audio signal restoration, while Chen et al. \cite{Chen2026} noted its impact on radar system angle estimation. Furthermore, Zhang et al. \cite{Zhang2025} investigated the effects of impulsive disturbances encountered during the transit of ocean-going vessels. Finally, Kumar et al. \cite{Kumar2023} highlighted the significance of impulsivity in the rail sector, establishing that train horns generate high-frequency impulsive acoustic signals that significantly disrupt the well-being of residents situated near railway infrastructure.

\subsection{Problem of non-Gaussian signals in 
 condition monitoring}\label{inne1}
Capturing physical measurements inevitably involves dealing with measurement errors and various interfering phenomena. This challenge is central to machine condition monitoring, which primarily utilizes vibration and acoustic data. Typically, the recorded signals are composed of two distinct parts: the informative component of interest and the non-informative noise \cite{randall2001relationship}. While many analytical models simplify this interference by assuming it follows a Gaussian distribution \cite{randall2001relationship}, real-world conditions often present disturbances that are non-Gaussian and distinctly impulsive \cite{Yu2013155}.

Such impulsive behavior significantly hinders the effective detection of machine defects \cite{wodecki2021,Zulawinski2024}, particularly when trying to identify local faults that manifest as periodic, impulsive signals. Consequently, understanding the specific characteristics of these disturbances is vital for optimizing the selection and efficacy of diagnostic algorithms.

The study of non-Gaussian impulsive noise paired with informative signals is an infrequent subject in existing condition monitoring literature. Nevertheless, fault identification within impulsive, non-Gaussian environments has gained considerable traction lately. Research has shown that various industrial processes—such as crushing \cite{Wyłomańska20165612}, cutting, packaging \cite{Cocconcelli2012667}, compression \cite{Barszcz2013473}, sieving \cite{Duda-Mroz2021}, and milling \cite{Kharchenko201725}—frequently produce impulsive components even during normal, healthy machine operation. Beyond mechanical processes, these disturbances can also arise from hardware malfunctions like sensor failures \cite{Jabłoński2013248}, data transmission issues, electromagnetic interference \cite{Mauricio2020,Smith2016371}, or control system artifacts \cite{Wodecki2025597}.

Because traditional diagnostic methods often fail when faced with the non-Gaussian nature of mechanical signals, recent literature has introduced various techniques designed to handle heavy-tailed, impulsive background noise. An early contribution to this field is found in \cite{mssp1}, which utilized $\alpha$-stable distributions for detecting faults in rolling element bearings. Further research has scrutinized how impulsive noise affects standard diagnostic tools; for instance, \cite{Borghesani2017378} examined its impact on square envelope and cyclic modulation spectra, while Wodecki et al. \cite{wodecki2021} explored its influence on classical cyclostationary analysis. Alternatively, Zhao et al. \cite{Zhao2019} implemented a diagnostic framework centered on the cyclic correntropy function. Other robust advancements include the use of generalized Gaussian cyclostationarity for blind deconvolution \cite{PENG2023110351}, maximum likelihood ratio frameworks for indicator design \cite{ANTONI2019290}, and specialized multi-fault strategies for gearboxes \cite{HE2021108738}. Additionally, \cite{Yi20251564} introduced an adaptive cyclic content ratiogram as an evolution of the fast kurtogram, and \cite{Luan2016503} demonstrated the superiority of cyclic correntropy over fractional lower-order statistics when dealing with $\alpha$-stable noise.

Our contributions  in this field include an improved infogram \cite{hebda2022infogram}, conditional variance-based statistics for frequency band identification \cite{cvb}, and robust second-order estimators for cyclostationary analysis \cite{Zulawinski2024}. Additionally, {we} have employed spectrogram factorization for impulsive bearing environments \cite{Gabor20242944} and non-negative tensor factorization to overcome the challenges of overlapping fault signatures and non-Gaussian disturbances in spectral structures \cite{Michalak2025}.

It should be mentioned that, while the studies mentioned above incorporate resilience to impulsive noise, they generally treat it as a secondary factor during specific diagnostic steps rather than focusing on the noise's fundamental modeling and evaluation. The discussion related to the problem of evaluating the impulsive behavior of the diagnostic signals is presented in Section \ref{intro}, where we indicate research articles in this area. 

\subsection{Example non-Gaussian-based models for industrial data}\label{inne2}
Diverse applications utilize various non-Gaussian noise models. In power-line communications, $\alpha$-stable distributions effectively capture statistical properties \cite{Laguna-Sanchez20151863}, while broader communication contexts often employ Bernoulli-Gaussian or Markov-Gaussian models with memory \cite{Shongwe2015119}. In condition monitoring, researchers primarily utilize $\alpha$-stable models \cite{robust_coherent} and Gaussian mixtures \cite{Zulawinski2024}, though hybrid models, such as the Gaussian-Pareto mixture, have also been proposed for vibration signals \cite{7465788}. Identifying the optimal model requires analyzing the dependence structure of the data \cite{ZULAWINSKI2023109677} and residual distributions. Beyond $\alpha$-stable models, other leptokurtic distributions including Burr, Weibull, and Gamma, or various mixture models, offer extensive flexibility for fitting specific physical phenomena from a statistical perspective \cite{rweron}.
\subsection{Measures for assessment of impulsive behavior of the signals}\label{inne3}
In condition monitoring, impulsive components are linked to machinery faults, especially if they are cyclic. The literature provides various metrics for quantifying data impulsivity.  Although fundamental statistics like sample kurtosis and variance are frequently utilized in condition monitoring, they often prove insufficient for accurately characterizing impulsive noise. Logically, as impulse amplitudes grow, a measure of impulsivity should increase proportionally; similarly, a higher recurrence of these impulses should indicate a more impulsive signal profile. Nevertheless, certain established indicators do not always follow this intuitive behavior.

In the realm of statistics, evaluating impulsivity extends beyond single-value metrics to the study of the "distance" between a baseline distribution (typically Gaussian) and the empirical distribution observed in the data. These discrepancies are generally assessed using cumulative distribution functions (CDF) or probability density functions (PDF), with the latter referred to as divergence measures. Common examples of these methodologies include the Kolmogorov-Smirnov statistic, which is CDF-based \cite{ks_test}, and the Hellinger distance, which relies on the PDF \cite{grzesiek2021method}.

Moreover, impulsivity can be analyzed through the framework of statistical hypothesis testing. This process involves determining, at a specific significance level, if a calculated test statistic deviates from what would be expected for a reference Gaussian signal. As a result, many traditional normality tests serve as effective tools for detecting signal impulsivity. Widely recognized examples include the Shapiro-Wilk and Lilliefors tests \cite{yap2011comparisons}. Additional statistical strategies, which may be less intuitive but remain highly adaptable for assessing impulsivity, are explored in the following sections of this article.

Although many metrics for assessing impulsivity are known in the literature, there is still a lack of objective methods to evaluate the scenarios in which these measures are effective and yield the desired results. The methodology proposed in this article addresses these challenges and can be successfully applied across many areas where the problem of non-Gaussian (impulsive) data is a natural phenomenon.

\section{Systematizing measures of impulsivity: theoretical framework} \label{sec:measures}

In this section, various statistical measures used for the
characterization of impulsive components in signals are reviewed and
systematized. 
In this work, impulsivity is understood as the presence of high amplitude, short-time disturbances in the time domain. They may appear according to a given cycle (normal operation of the compressor) or can have a random location (pure disturbance).  Short-time high amplitude disturbances in the time domain can be represented as wideband spectrum disturbances in the frequency domain or as heavy-tailed behavior from the distribution point of view. In this work, we focus on the properties of signals in the time domain or distribution domain.
Impulsive components might manifest themselves in several ways: increased dispersion, strong local peaks, or deviations of
the empirical distribution from a chosen reference model. Consequently, numerous
measures have been proposed in the literature to quantify such
behavior. The following subsections summarize the most commonly used
approaches and group them according to their theoretical foundations.


In this work we distinguish five main categories of impulsivity measures - grouped according to the statistical principle on which they are based: basic statistics, combinations of basic statistics, estimated distribution parameters, distribution-distance measures, and statistical test measures. All of the considered measures are presented in Tables \ref{tab1}-\ref{tab5} given in \ref{app:tables}. It is important to notice that each mentioned group highlights a different aspect of impulsive signal structure. 


\subsection{Basic statistics} \label{sec:measures-group1}

The first group of measures consists of classical low-order statistics that describe the fundamental properties of the amplitude of the distribution. These measures quantify the spread, asymmetry, and tail behavior of the observed samples without assuming a specific impulsive model.

Standard deviation (SD) and root mean square (RMS) describe the overall amplitude variability and energy of the signal, respectively, whereas higher-order standardized moments such as skewness (SKEW) and kurtosis (KURT) provide additional information about the shape of the distribution. In particular, skewness characterizes asymmetry of the amplitude distribution, while kurtosis reflects the relative importance of extreme values and heavy
tails, which are commonly associated with impulsive disturbances.

The definitions of the considered basic statistics are summarized in
Table~\ref{tab1} in \ref{app:tables}.

\subsection{Combinations of basic statistics}

The second group consists of measures obtained by combining simpler statistics into scale-normalized ratios or inequality descriptors. Such measures aim to suppress the influence of the absolute signal scale and highlight structural features related to impulsive behavior. 

Typical examples include the shape factor (SF),
which relates quadratic and absolute averages of the signal, and the crest factor (CF), which compares the largest
instantaneous amplitude to the overall signal energy. Information-theoretic quantities such as entropy (ENT) and negentropy (NEG) provide
a complementary perspective by characterizing the disorder or non-Gaussian
structure of the amplitude distribution.
Finally, the Gini index (GINI) quantifies the degree of inequality in the magnitude
distribution and can therefore reflect the concentration of energy in a small
number of samples.

The considered measures of this category are summarized in
Table~\ref{tab2} in \ref{app:tables}.

\subsection{Estimated distribution parameters}

The third group of impulsivity descriptors is based on parameters of fitted probabilistic models. Instead of relying only on empirical moments, these measures characterize impulsive behavior through parameters of heavy–tailed distributions estimated from the observed data. In an impulsive setting, the distributions often exhibit heavier
tails than the Gaussian model. Such behavior can be modeled using parametric families that explicitly control the tail decay. Typical examples include the $\alpha$-stable distribution \cite{samorodnitsky1994stable}, Student's $t$ distribution \cite{gosset1908probable}, and the generalized Gaussian distribution \cite{nadarajah2005generalized}.

In the $\alpha$-stable model, the stability parameter $\alpha \in (0,2]$ controls the heaviness of the distribution tails.
Smaller values of $\alpha$ correspond to heavier tails and therefore stronger impulsive behavior. For convenience, the quantity $2-\alpha$ is sometimes used as a direct
indicator of impulsivity, since it increases as the distribution departs from the Gaussian case ($\alpha=2$). This measure is denoted here as 2-ALPHA. Similarly, the scale parameter $\sigma$ of the $\alpha$-stable model quantifies the overall intensity of fluctuations and provides a robust measure of signal variability, even when classical second-order moments do not exist. In this article, we denote this measure as SIGMA.  Another commonly used model is the Student's $t$ distribution,
characterized by the number of degrees of freedom parameter $\nu$ (denoted later as TDOF - from Student's t  number of degrees of freedom). Small values of $\nu$ correspond to heavy-tailed distributions and frequent extreme values, whereas large values of $\nu$ approach the Gaussian distribution. Finally, the generalized Gaussian distribution introduces a shape parameter $\beta$ controlling the decay rate of the distribution tails. When $\beta<2$, the distribution becomes more peaked and heavy-tailed than the Gaussian case.
For convenience, the quantity $2-\beta$ can again be interpreted as a direct indicator of deviation from Gaussian behavior. In this article, we note this measure as 2-BETA.

These parameter-based measures provide a compact way of describing impulsive characteristics through interpretable model parameters. The considered measures and their estimation methods are summarized in
Table~\ref{tab3} in \ref{app:tables}.

\subsection{Distribution-distance measures}

The fourth category includes measures that quantify the discrepancy between the empirical distribution of the observed signal and a chosen reference distribution - which can be understood as the distribution of the non-impulsive background signal. Unlike parameter-based methods, these measures do not rely on a specific parametric model but instead directly compare probability distributions. 
Distribution-distance measures quantify the difference between the probability distributions in various mathematical senses. 

We differentiate two groups of those statistics: cumulative
distribution function-based (CDF-based) and probability density function-based (PDF-based). In the first group, we consider the Kolmogorov-Smirnov statistic (KS) - that measures the maximum vertical distance between the empirical and reference CDFs, and the Cramér-von Mises statistic (CvM) - that evaluates the integrated squared difference between the empirical and theoretical CDFs. The CvM statistic provides a more global measure of the distributional discrepancy, highlighting the differences more in the tails. In the second group of measures, metrics based on probability densities, we consider the Hellinger divergence (HD) that measures the similarity between two distributions by comparing the square-root transformed densities, and the Jeffreys divergence (JD) which is a symmetrized version of the Kullback-Leibler divergence, measures the discrepancy between the empirical and reference probability density functions in terms of information content. 

In the context of impulsive signals, large values of these measures indicate strong deviations of the empirical distribution from the assumed background model, which may be caused by heavy-tailed amplitude distributions or rare extreme events. The considered distribution-distance measures are summarized in
Table~\ref{tab4} in \ref{app:tables}.

\subsection{Test statistics}
The final category consists of statistics used in formal goodness-of-fit hypothesis testing procedures, i.e., the test statistics. Let us mention that the statistical test returns binary information (yes or no), while the test statistic may take any value  depending on its definition. Test statistics in goodness-of-fit testing evaluate whether the observed signal is statistically consistent with a chosen reference distribution. Typically, the null hypothesis assumes that the signal samples follow a specific background distribution (often Gaussian). Large values of the corresponding test statistics indicate deviations from this assumption. 

The Jarque-Bera (JB) measure combines skewness and kurtosis into a single test statistic that evaluates whether the third and fourth standardized moments match those of the Gaussian distribution. It is particularly sensitive to deviations caused by heavy tails or asymmetric impulsive disturbances. The Shapiro-Wilk (SW) statistic evaluates normality by analyzing the
correlation between ordered sample values and expected order
statistics of a Gaussian distribution. This test is known for its high sensitivity to deviations in both skewness and kurtosis. The Epps-Singleton (ES) statistic compares empirical and theoretical characteristic functions, and it can detect a broad class of deviations from the assumed Gaussian model. Finally, the modified Greenwood (MGS) statistic measures the unevenness of energy distribution within the signal. In impulsive signals, energy is often concentrated in a small number of samples, which leads to larger values of this statistic. The considered test-based measures are summarized in
Table~\ref{tab5} in \ref{app:tables}.

\subsection{Summary}

The taxonomy presented in the above subsections highlights that the impulsivity of a given signal is a property that cannot be fully captured by a single descriptor. Different measures emphasize different statistical aspects of impulsive behavior and therefore may respond differently depending on the characteristics of the underlying background signal. In diagnostic applications, analyzed signals may exhibit various structures, including deterministic patterns or correlated dynamics. As a consequence, the effectiveness of individual impulsivity measures depends not only on the presence of impulsive disturbances but also on the statistical properties of the background signal itself. Another important difficulty arises from the fact that these measures are defined on different scales and have different statistical interpretations. Direct comparison of their numerical values is therefore generally not meaningful. Instead, their relative effectiveness must be evaluated through systematic analysis, for example by studying their behavior under controlled signal models or simulation scenarios.



\section{Methodological approach to impulsivity estimation: selection criteria and statistical evaluation} \label{sec:methodology}


The previous section introduces a taxonomy of impulsivity measures based on their statistical properties and theoretical foundations. However, the existence of multiple descriptors raises a question on how different impulsivity measures can be compared and evaluated in practice. Because impulsivity measures are defined on different mathematical scales and often capture different statistical aspects of the signal, their numerical values are not directly comparable. Moreover, their sensitivity may depend on the structure of the background signal, the amplitude of impulsive components, and the frequency of their occurrence. For these reasons, a consistent evaluation framework is required in order to assess the ability of different measures to detect or quantify impulsive behavior. 

In this work, the comparison of impulsivity criteria is formulated as a statistical discrimination problem between two classes of signals: a reference signal representing the background component and a test signal containing certain disturbances. In the considered case, the difference between these classes is associated with impulsive behavior, although the same framework can be applied to any situation in which the two signals differ in their statistical properties.

\begin{figure}[h!]
    \centering
    \includegraphics[width=\textwidth]{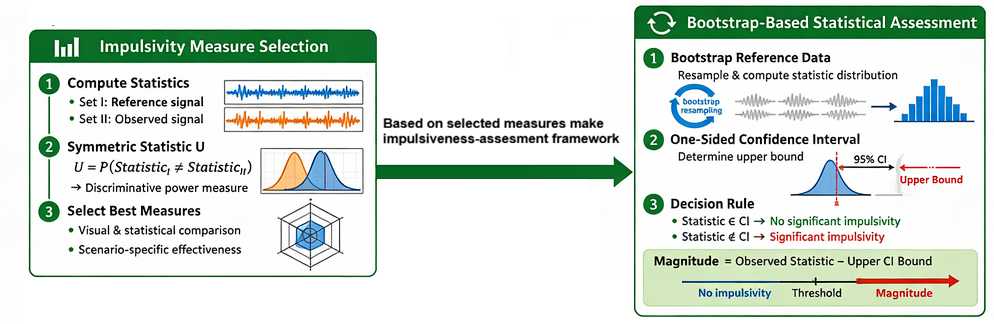}
    \caption{The general framework for impulsivity assessment.}
\label{fig:framework}
\end{figure}

The proposed methodology consists of a two-stage procedure, illustrated in Fig.~ \ref{fig:framework}. In the first stage, the considered impulsivity measures are evaluated in terms of their ability to discriminate between the reference and test signals. This step identifies measures that provide the strongest statistical separation between the two classes in a given scenario. To perform this comparison, a statistic with a simple probabilistic interpretation is used, which quantifies the probability that the value of a given measure computed for the test signal exceeds the corresponding value obtained for the reference signal. In the second stage, the measures with the strongest discrimination ability are used in a bootstrap-based statistical assessment framework to determine whether the observed value of a given measure indicates statistically significant impulsivity and to quantify its magnitude.

The following subsections describe the proposed methodology for comparing impulsivity criteria and the statistical procedure used for assessing their effectiveness.

\subsection{Impulsivity criteria selection}
\label{sec:criteria_selection}

In this section, the comparison of different impulsivity criteria is formulated as a statistical discrimination problem between two
classes of signals: the class of the reference
signal (\textit{ref}), and the class of the test
signal (\textit{test}). Let $\mathbf{x}:= x_1, x_2, \ldots, x_k$ denote the reference signal that is the realization of a random process $X^{(ref)}$, let $\mathbf{y}:=y_1, y_2, \ldots, y_k$ denote the test signal (with a possible impulsive component with respect to the reference signal) that is the realization of a random process $Y^{(test)}$, and let $M(\cdot)$ denote an impulsivity measure applied to that signal. It might be any of the measures considered in Section \ref{sec:measures} or other criteria not discussed in this article. The value of the criterion is computed as
\begin{equation}
m^{(ref)} = M(x_1, x_2, \ldots, x_k)
\label{eq:m_ref}
\end{equation}
for the reference signal, and
\begin{equation}
m^{(test)} = M(y_1, y_2, \ldots, y_k)
\label{eq:m_obs}
\end{equation}
for the test signal. To evaluate the behavior of a given measure, we consider $N$ realizations of
both signal classes. Let
$$\{m^{(ref)}_1, m^{(ref)}_2, \ldots, m^{(ref)}_{N}\}$$
denote the values of the measure computed for $N$ realizations of the reference signal and let
$$\{m^{(test)}_1, m^{(test)}_2, \ldots, m^{(test)}_{N}\}$$
the corresponding values obtained for $N$ realizations of the test
signal containing impulsive components. 

The ability of a given criterion to distinguish between
these two signal classes can be quantified by measuring the degree of
separation between the distributions of $m^{(ref)}$ and $m^{(test)}$. Here, this separation is evaluated using the Mann-Whitney
statistic \cite{mann1947test}, which provides a nonparametric measure
of stochastic distinction between two random variables. In this work, we express the Mann-Whitney statistic in the language of measure $M(\cdot)$ realizations as
\begin{equation}
U =
\sum_{i=1}^{N}
\sum_{j=1}^{N}
\left[
I(m^{(test)}_i > m^{(ref)}_j)
+
\frac{1}{2}
I(m^{(test)}_i = m^{(ref)}_j)
\right],
\end{equation}
where $I(A)$ denotes the indicator function defined as
\begin{equation}
I(A) =
\begin{cases}
1, & A \text{ is true}, \\
0, & A \text{ is false}.
\end{cases}
\end{equation}
Let us notice that the statistic counts the number of pairwise comparisons in which the value of the measure obtained for the test signal exceeds the value
obtained for the reference signal, while ties contribute one half. For comparison of impulsivity measures a normalized form of this
statistic is used
\begin{equation} \label{eq:U_norm}
U_{\mathrm{norm}} =
\frac{U}{N^2},
\end{equation}
which takes the values in the interval $[0,1]$.
Moreover, the normalized statistic given in Eq. (\ref{eq:U_norm}) aligns with a probabilistic
interpretation of the following form
\begin{equation}
U_{\mathrm{norm}}
=
P(M(Y^{(test)}) > M(X^{(ref)}))
+
\frac{1}{2}
P(M(Y^{(test)}) = M(X^{(ref)})).
\end{equation}
Thus $U_{\mathrm{norm}}$ represents the probability that a randomly
selected realization of the test signal produces a larger value of
the impulsivity measure than a randomly selected realization of the
reference signal. Values of $U_{\mathrm{norm}}$ close to $0.5$ indicate that the
distributions of the measure for the two signal classes strongly
overlap, meaning that the considered criterion has no ability
to detect impulsive disturbances - it does not differente the signal classes. In contrast, values approaching $0$ or $1$ indicate strong separation of the
distributions and therefore high discriminative capability of the
measure.

It should be noted that the normalized Mann-Whitney statistic is
mathematically equivalent to the area under the receiver operating
characteristic curve (AUC), commonly used in
classification problems \cite{hanley1982meaning,bradley1997use}.
Both quantities represent the probability that a randomly chosen
observation from one class is ranked higher than a randomly chosen
observation from the other class. Therefore, $U_{\mathrm{norm}}$ provides a natural, scale-invariant
measure of the effectiveness of a given criterion in
distinguishing between two signals. However, because different impulsivity measures considered in Section \ref{sec:measures} may either increase or decrease
in the presence of impulsive components, the final comparison criterion
is defined as follows
\begin{equation}
U^* = \max(U_{\mathrm{norm}}, 1 - U_{\mathrm{norm}}),
\end{equation}
which ensures that the resulting score always reflects the separation
capability of the measure independently of the direction of its
response.
Consequently, $U^*$ lies in the interval $[0.5,1]$, where larger
values correspond to better discrimination between reference and
impulsive signals.

In practical implementation, the above quantities are estimated
empirically. For each impulsivity measure the statistics $m^{(ref)}$ and
$m^{(test)}$ are obtained from multiple realizations of the considered
signal models. The resulting empirical distributions are then compared using the
statistic defined above, enabling systematic comparison of different
impulsivity criteria under identical signal conditions. Multiple realizations of the signals may be generated through controlled simulation
scenarios - see Section \ref{sec:mc_sim}, or extracted from experimental datasets - for details we refer to the real data analysis in Section \ref{sec:real}. 

\subsection{Bootstrap-based impulsivity assessment}\label{BIAA}
As a second step of the proposed framework, the selected impulsivity criteria (received using the procedure presented in Section \ref{sec:criteria_selection}) are further analyzed. We introduce here  a \emph{bootstrap-based impulsiveness assessment} (BIA) procedure designed to determine whether the test signal (\textit{test}) exhibits a statistically significant change in impulsiveness relative to the reference signal (\textit{ref}). Additionally, the method quantifies the magnitude of this change using the \emph{impulsiveness magnitude factor} (IMF), which is defined later in this section.

For consistency, we use the same notation as in Section~\ref{sec:criteria_selection}. The BIA framework assumes that, for the selected impulsiveness criterion $M(\cdot)$, a statistic of the following general form is evaluated
\begin{align}
\mathcal{M}(M, \mathbf{x}, \mathbf{y}) := \left|M(x_1, x_2, \ldots, x_k) - M(y_1, y_2, \ldots, y_k)\right|.
\label{eq:stat1}
\end{align}

The transformation defined in~\eqref{eq:stat1} is introduced to obtain a unified and easily interpretable measure, independent of the specific choice of the statistic $M(\cdot)$. In particular, $\mathcal{M}(M)$ quantifies the absolute difference between the impulsiveness measure computed for the reference and test signals. As a result, $\mathcal{M}(M)$ takes values in the range $[0, \infty)$, where values close to zero indicate that a given measure does not indicate a significant difference between reference and tested signal (i.e. their impulsiveness is comparable) while larger values correspond to increasing discrepancies. 

To determine the expected range of $\mathcal{M}(M)$ under the null hypothesis, i.e., no change in impulsiveness, a bootstrap procedure is employed. Specifically, subsamples are repeatedly drawn from the reference signal to approximate the distribution of $\mathcal{M}(M)$ under the assumption that both signals originate from the same process. This procedure is described in detail below, while  in \ref{app:bootstrap_app},  the version of this algorithm for dependent data (used in simulated scenarios in Section \ref{sec:mc_sim}) is discussed. Moreover, the formulation in~\eqref{eq:stat1} facilitates a natural definition of the IMF, as discussed in the subsequent part of the algorithm. 

The detailed steps of the BIA procedure are summarized below:

\begin{enumerate}

\item \textbf{Definition of signals and statistic}\\
Now, we assume that the length of the reference signal $\mathbf{x}$ is equal to $s$ while the length of the test signal $\mathbf{y}$ is equal to $k$, where $s \gg k$. The condition $s \gg n$ is required to enable reliable bootstrap resampling from the reference signal. Let $M(\cdot)$ denote a selected impulsivity criterion. First, the difference in impulsiveness between the reference and test signals is quantified using Eq.~\eqref{eq:stat1}.

\item \textbf{Bootstrap reference distribution}

To approximate the distribution of $\mathcal{M}(M)$ under the null hypothesis of no change in impulsiveness, we  generate $r$ bootstrap realizations using  the reference signal. In each $l$-th iteration:
\begin{itemize}
\item draw two independent subsamples $\mathbf{x}^{(1)}, \mathbf{x}^{(2)}$ of length $k$ from $x_1, x_2, \ldots, x_s$,
\item compute the statistic $\mathcal{M}^{(k)}:=\mathcal{M}(M, \mathbf{x}^{(1)}, \mathbf{x}^{(2)})$ for the obtained pair.
\end{itemize}
The resulting values form an empirical distribution of $\mathcal{M}(M, \mathbf{x}^{(1)}, \mathbf{x}^{(2)})$ under the assumption of homogeneity, i.e., assuming that both samples come from the same model (and have the same parameters). In other words, we compute the value of the statistic under the assumption that the data from the two samples share the same characteristics.

\item \textbf{Confidence interval construction}

Based on the bootstrap distribution, construct a one-sided confidence interval at confidence level $c$
\begin{eqnarray}
\mathcal{R}_c := (-\infty, \mathcal{M}_c],
\end{eqnarray}
where $\mathcal{M}_c$ denotes the corresponding  quantile of order $c$ calculated based on the empirical distribution of $\mathcal{M}(M, \mathbf{x}^{(1)}, \mathbf{x}^{(2)})$.

\item \textbf{Evaluation of the observed statistic}

Compute the observed value
\[
\mathcal{M}_{\mathrm{obs}} = \mathcal{M}(M, \mathbf{x}, \mathbf{y}),
\]
using the reference signal (restricted to $n$ last samples) $\mathbf{x}$, and the test signal $\mathbf{y}$. 

\item \textbf{Decision rule}

The test signal is classified as exhibiting significantly different impulsiveness relative to the reference signal if $\mathcal{M}_{\mathrm{obs}} \notin \mathcal{R}_c$. Otherwise, no statistically significant difference in impulsiveness is detected. 

\item \textbf{Impulsiveness magnitude factor}

If a statistically significant difference is detected, its magnitude is quantified by the IMF defined as follows
\begin{align}
\mathrm{IMF} = \frac{\mathcal{M}_{\mathrm{obs}} - \mathcal{M}_c}{\mathcal{M}_c}.
\end{align}

\end{enumerate}

\section{Analysis of simulated signals from the adopted models}
\label{sec:mc_sim}
\subsection{Scenarios definition}

To demonstrate the efficiency of the methodology presented in Section \ref{sec:methodology} under controlled conditions, three synthetic signal scenarios are constructed. All scenarios represent signals generated under normal operating conditions of a mechanical system. The purpose of these simulations is to reproduce selected signal characteristics observed in healthy machines, including stochastic dynamics of the signal and additional deterministic (periodic) or random impulsive components.

The baseline signal is modelled as a stationary autoregressive process of order $p$, denoted as AR($p$). The signal is defined as \cite{brockwell2016introduction}
\begin{equation}
X_t = \sum_{k=1}^{p} \theta_k X_{t-k} + \varepsilon_t ,
\end{equation}
where $X_t$ denotes the value at discrete time $t$, $\theta_k$ are the autoregressive coefficients, and $\varepsilon_t$ is a white Gaussian noise process, i.e. it is a sequence of uncorrelated random variables such that for each $t$, $
\varepsilon_t \sim \mathcal{N}(0,\sigma^2)$, where $\sigma^2>0$ denotes the variance. The coefficients $\theta_k$ are selected such that the process is stationary. The AR($p$) model captures the temporal correlation structure commonly observed in the signals generated by mechanical systems, however the used model is the simplified version of the model observed in the real signals. In Table \ref{tab:simulation_scenarios} we specify the details of the simulation scenarios, and below we describe them in detail. In Table \ref{tab:symbols} we listed all symbols used in the simulated scenarios description.

\begin{table}[h!]
\centering
\caption{Definition of the simulated signal scenarios considered in the study.}
\label{tab:simulation_scenarios}
\begin{tabular}{lll}
\hline
Scenario & Mathematical model & Description \\
\hline

Scenario 1 &
$S_t^{(1)} = X_t$ &
Baseline signal generated by an AR($p$) process. \\

Scenario 2 &
$S_t^{(2)} = X_t + i_t$ &
AR($p$) baseline with periodic impulsive events. \\

Scenario 3 &
$S_t^{(3)} = X_t + b_t$ &
AR($p$) baseline with periodic damped oscillatory responses. \\

\hline
\end{tabular}
\end{table}
\begin{table}[h!]
\centering
\caption{Symbols used in the simulated scenarios description}
\label{tab:symbols}
\begin{tabular}{ll}
\hline
Symbol & Description \\
\hline
$X_t$ & stochastic baseline modelled as an AR($p$) process \\
$i_t$ & periodic impulse sequence with amplitude $I$ and period $T_p$ \\
$b_t$ & damped oscillatory burst generated by an excitation \\
$I$ & impulse amplitude \\
$T_p$ & excitation period in samples \\
$f_s$ & sampling frequency \\
$f_0$ & oscillation frequency of the burst \\
$\tau_d$ & exponential decay constant of the oscillatory response (in seconds) \\
$\phi$ & initial phase of the oscillation \\
\hline
\end{tabular}

\end{table}

\paragraph{Scenario 1: Baseline Stochastic Model}

In Scenario 1, the signal only consists of the stochastic AR($p$) process, i.e.,
\begin{equation}
S_t^{(1)} = X_t.
\end{equation}
This case represents the baseline response of a healthy mechanical system without additional disturbances.

\paragraph{Scenario 2: Simplified Compressor Model}

In Scenario 2 periodic impulsive events are added to the stochastic baseline, i.e.,
\begin{equation}
S_t^{(2)} = X_t + i_t .
\end{equation}
The impulsive periodic component is defined as a sequence of short-duration impulses
\begin{equation}
i_t =
\begin{cases}
I, & t = kT_p, \quad k \in \mathbb{Z}_{\ge 0}, \\
0, & \text{otherwise},
\end{cases}
\end{equation}
where $I$ denotes the impulse amplitude and $T_p$ is the impulse period in samples. From a theoretical perspective, a single impulse can be represented by the Dirac delta function $\delta(t - t_0)$, where $t_0$ denotes the time of occurrence. In the discrete-time setting considered here, it is approximated by a single-sample impulse. In the context of the real data analyzed in Section \ref{sec:real}, this scenario can be interpreted as a simplified model of a compressor operating under normal conditions, where, in addition to the stochastic dynamics of the signal, periodic impulsive interactions between mechanical components occur. The adopted formulation uses a highly simplified representation of these events as one-sample outliers.

\paragraph{Scenario 3: Extended Compressor Model.} In Scenario 3 the periodic disturbances generate a short oscillatory response rather than a single impulse. The resulting signal is then defined as 
\begin{equation}
S_t^{(3)} = X_t + b_t.
\end{equation}
Each disturbance produces a decaying oscillatory burst described by
\begin{equation}
\tilde{b}(t) = I\,e^{-t / (\tau_d f_s)} \sin \left( 2 \pi \frac{f_0}{f_s} t + \phi \right),
\end{equation}
where $t$ denotes the discrete time index within the oscillatory burst, $I$ is the amplitude, $f_0$ is the oscillation frequency, $\tau_d$ is the exponential decay constant expressed in seconds, $f_s$ is the sampling frequency, and $\phi$ is the initial phase. The periodic burst signal is therefore constructed as
\begin{equation}
b_t = \sum_{k=0}^{\infty} \tilde{b}(t-kT_p).
\end{equation}
This model more closely reflects the behavior of real compressor systems, see Section \ref{sec:real}, where periodic components do not manifest as one-sample impulses but rather as responses of the mechanical structure. In practice, such events excite resonant modes, resulting in short oscillatory waveforms resembling damped harmonic responses.

\begin{figure}[h!]
    \centering
    \includegraphics[width=0.7\textwidth]{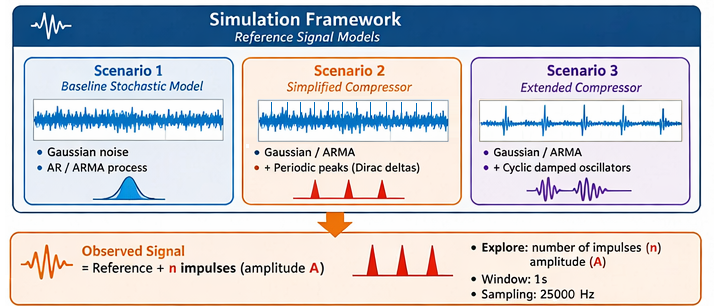}
    \caption{The reference signals under adopted  scenarios.}
    \label{fig:scenarios}
\end{figure}

To reflect the dynamics of real mechanical signals, we extend the simulated data by introducing an additional impulsive component, and the impulsiveness assessment is provided concerning this component. The impulses are modelled as single-sample observations with a prescribed amplitude, superimposed at random time instants on the stochastic signal. Since the methodology proposed in Section \ref{sec:measures} consists of selecting measures of impulsiveness, followed by statistical testing and magnitude assessment, the simulation framework explicitly incorporates different levels of impulsiveness. This is achieved by considering signals with varying numbers of impulses, controlled by the parameter $n$, as well as different impulse amplitudes $A$. In this way, both the frequency and the amplitude of impulsive events, which determine the level of impulsiveness of the signal, can be taken into account when evaluating the efficiency of the methodology. The general concept of the simulation framework and the considered signal models is illustrated in Fig.~ \ref{fig:scenarios}. 

The parameters of the simulation study are summarized in Table \ref{tab:simulation_parameters}. All signals were generated $MC=1000$ times with a sampling frequency of $f_s = 25000$~Hz and a duration of $1$s, using a stationary AR(2) process as stochastic baseline. Scenario-specific periodic components and additional impulsive events were defined according to the specifications provided in the table. In the simulation study, the signals generated under Scenarios 1-3 without the additional impulsive component are considered as reference signals, whereas those with added impulses are treated as test signals, in accordance with the nomenclature introduced in Section \ref{sec:methodology}.

\begin{table}[ht]
\centering
\caption{Parameters used in the simulation study.}
\label{tab:simulation_parameters}
\begin{tabular}{lll}
\hline
Component & Parameter & Value \\
\hline
General & Sampling frequency $f_s$ & $25\,000$ Hz \\
General & Signal duration & $1$ s \\
General & Number of samples & $25\,000$ \\

Baseline AR(2) & $\theta_1$ & $0.6$ \\
Baseline AR(2) & $\theta_2$ & $-0.2$ \\
Baseline AR(2) & Noise standard deviation $\sigma$ & $1.0$ \\
Baseline AR(2) & Theoretical standard deviation of $X_t$ & $\approx 1.18$ \\

Scenario 2 & Period $T_p$ & $2\,500$ samples \\
Scenario 2 & Impulse amplitude $I$ & $8.0$ \\

Scenario 3 & Period $T_p$ & $2\,500$ samples \\
Scenario 3 & Burst amplitude $I$ & $8.0$ \\
Scenario 3 & Oscillation frequency $f_0$ & $4000$ Hz \\
Scenario 3 & Decay constant $\tau_d$ & $0.005$ s \\
Scenario 3 & Initial phase $\phi$ & $0$ rad \\

Additional impulsive component & Number of impulses $n$ & $0,1,\ldots, 24, 25$ \\
Additional impulsive component & Impulse amplitude $A$ & $5, 5.5, \ldots, 11.5, 12$ \\
\hline
\end{tabular}
\end{table}

\subsection{Impulsivity criteria selection -- results for Scenario 1}
\label{sec:imp_crit_s1}
\begin{figure}[h!]
    \centering
    \includegraphics[width=0.8\textwidth]{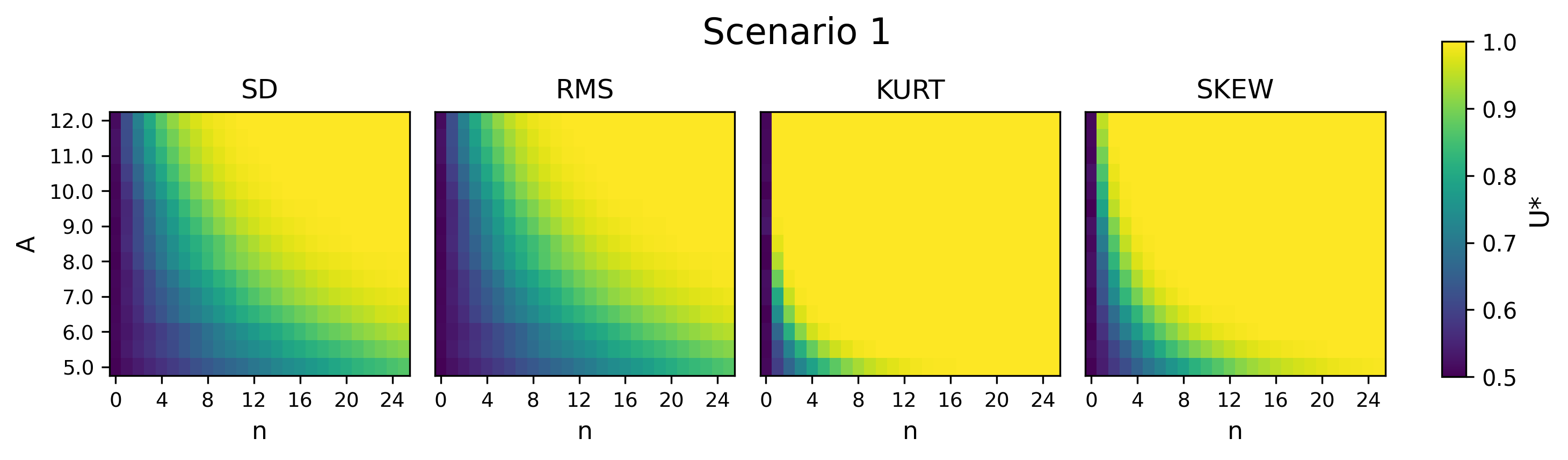}
    \caption{Comparison of $U^*$ heatmaps for Scenario 1 for the impulsivity metrics presented in Tables \ref{tab1}.}
\label{fig:heatmap1}
\end{figure}

The results for Scenario~1, corresponding to the first group of statistics described in Section \ref{sec:measures-group1}, are presented in Fig.~ \ref{fig:heatmap1}. The values of the $U^*$ statistic are visualized using heatmaps as a function of the number of generated impulses $n$ and their amplitude $A$. 

As expected, for $n = 0$, the test signal is equivalent to the reference signal, resulting in $U^* \approx 0.5$ for all considered statistics. This value indicates that the distributions of the statistics for the reference and test signals are indistinguishable. As both the number of impulses $n$ and their amplitude $A$ increase, the values of $U^*$ systematically rise, reflecting an increasing divergence between the distributions of the analyzed statistics. This confirms that all considered measures are sensitive to the presence of impulsive components, although the degree of sensitivity varies between them.

Among the metrics from this group, SD and RMS exhibit a gradual and smooth increase of $U^*$, indicating consistent sensitivity to the overall energy increase caused by impulsive events. However, the transition between low and high discrimination regions is relatively diffuse, suggesting limited sensitivity to lower-level impulsive disturbances. In contrast, SKEW shows a more pronounced increase in $U^*$, particularly for lower values of $n$ combined with higher amplitudes. This indicates higher sensitivity to asymmetry introduced by impulsive components. The most distinctive behavior is observed for KURT, which achieves high $U^*$ values already for relatively small numbers of impulses and moderate amplitudes. This confirms its strong sensitivity to impulsive and heavy-tailed behavior in this scenario, making it particularly suitable for early detection of impulsive disturbances. 

To quantify the sensitivity of each statistic to impulsive disturbances, a set of aggregate performance metrics based on the $U^*$ statistic was defined. For each combination of scenario and statistic, the same set of measures is evaluated under identical parameter settings:
\begin{itemize}
    \item \textbf{Impulse-region mean}: the average $U^*$ computed only for $n > 0$, reflecting performance when impulsive components are present. This metric quantifies the overall sensitivity of a statistic to impulsive disturbances, excluding the trivial case of identical signals ($n=0$).
    \item \textbf{Hard-region mean}: the average $U^*$ for low-amplitude impulses ($A \leq A_{\max}$), representing detection performance in challenging conditions. Here, $A_{\max} = 8$, which corresponds to twice the amplitude of the reference signal. This measure is particularly important for early fault detection, as it evaluates the ability of a statistic to distinguish lower-level impulsive changes from the reference signal.
    \item \textbf{Easy-region mean}: the average $U^*$ for higher amplitudes ($A > A_{\max}$), corresponding to clearly distinguishable impulsive cases. Here, $A_{\max} = 8$. It reflects the maximum achievable discrimination performance in the presence of strong impulsive disturbances.
    \item \textbf{Coverage}: the fraction of the parameter space where $U^*$ exceeds a predefined threshold (here $0.8$), indicating how often a given statistic provides strong discrimination.
    \item \textbf{Worst-case performance}: the minimum $U^*$ value observed for $n > 0$, representing robustness under the least favorable conditions.
\end{itemize}

\begin{table}[htbp]
\centering
\caption{$U^*$-based performance metrics for Scenario~1 for the impulsivity metrics presented in Table \ref{tab1}.} 
\label{tab:u_star_summary}
\begin{tabular}{l c c c c c}
\hline
Metric & Mean imp. & Mean hard & Mean easy & Coverage & Worst case \\
\hline
KURT & \textbf{0.988} & \textbf{0.975} & \textbf{1.000} & \textbf{0.941} & \textbf{0.593} \\
SKEW & 0.963 & 0.931 & 0.991 & 0.890 & 0.541 \\
SD & 0.866 & 0.799 & 0.925 & 0.674 & 0.523 \\
RMS & 0.866 & 0.799 & 0.925 & 0.674 & 0.523 \\

\hline
\end{tabular}
\end{table}

Table~\ref{tab:u_star_summary} summarizes the performance of the considered statistics for Scenario 1. The reported values correspond to the aggregate metrics defined above. The results show that KURT achieves the highest scores, followed by skewness, while energy-based measures (SD and RMS) exhibit noticeably lower performance. An interesting insight can be obtained by comparing \textit{Mean hard} and \textit{Mean easy}. While all statistics achieve near-perfect discrimination in the easy region (high-amplitude impulses), significant differences emerge in the hard region. In particular, KURT and SKEW maintain relatively high performance even for low-amplitude impulses, whereas SD and RMS exhibit a significant drop in discriminative capability. Overall, the results for Scenario 1 confirm that higher-order statistics (KURT and SKEW) provide superior sensitivity to impulsive disturbances, especially in challenging detection conditions, while energy-based measures are less effective for early-stage impulsive fault detection.

In Fig.~ \ref{fig:heatmap4x4_scenario1} in \ref{app:figures} one can see the analogous heatmaps for the remaining statistics presented in Section \ref{sec:measures} for Scenario 1. The comparison reveals distinct groups of statistical behavior with respect to impulsive disturbance sensitivity. Several metrics, such as CF, ENT, JB, SW, exhibit rapid increase to high $U^*$ values even for relatively low numbers of impulses or moderate amplitudes, indicating strong sensitivity for early detection tasks. In contrast, metrics such as SF, 2-ALPHA, TDOF, 2-BETA, and MGS exhibit a more gradual and smooth transition across the $(A,n)$ space, providing a wider dynamic range and thus greater potential for quantifying the severity of impulsive behavior, while still maintaining high $U^*$ values, indicating good overall discriminative performance. Other measures, including GINI, SIGMA, KS and CvM, demonstrate consistently low $U^*$ values, indicating poor discriminative capability in this scenario. Metrics such as HD and JD exhibit a distinct behavior, with a strong dependence on the number of impulses rather than their amplitude. In particular, HD shows a sharp transition once the number of impulses exceeds a certain threshold, even for relatively low amplitudes. This suggests that these metrics are more sensitive to the frequency of impulsive events than to their magnitude. The aggregated metrics for statistics considered in Fig.~ \ref{fig:heatmap4x4_scenario1} are presented in Table \ref{tab:u_star_all} in in \ref{app:figures}. Overall, the results confirm that higher-order and PDF-based statistics significantly outperform both energy-based and CDF-based measures in detecting impulsive behavior in Scenario 1.

It is worth to emphasize that metrics for which the values of $U^*$ rapidly icrease to $1$ can be interpreted as highly sensitive detectors of impulsive behavior. Such statistics are potentially well suited for fast alarm triggering, as they respond strongly even to weak impulsive components. However, this rapid increase may also indicate limited discriminative resolution at higher levels of impulsivity, i.e., different combinations of $n$ and $A$ might yield to similarly high $U^*$ values, making it difficult to distinguish between varying degrees of impulsive severity. In contrast, statistics characterized by a more gradient-like transition might provide a broader dynamic range of $U^*$ values. Although they may be less sensitive to weak impulsive events, they might offer improved capability for differentiating between different levels of impulsivity. This property is particularly important in applications where not only detection, but also quantification or monitoring of fault progression is required. Therefore, from a practical perspective, the choice of statistic depends on the intended application. Therefore, the second step of the procedure described Section \ref{sec:methodology} is essential, as it enables the assessment of which statistic is most suitable for quantifying the magnitude of impulsive disturbances, for details of this analysis see Section \ref{sec:impulsivity_assessment}.

\subsection{Impulsivity criteria selection -- results for Scenario 2}

\begin{figure}[h!]
    \centering
    \includegraphics[width=0.8\textwidth]{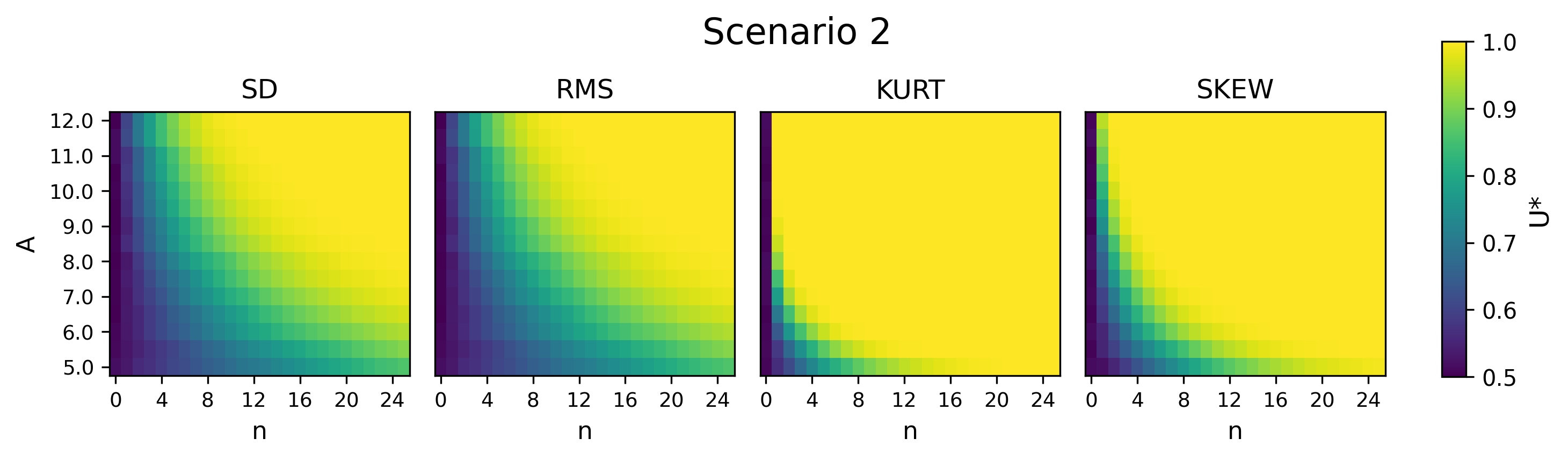}
    \caption{Comparison of $U^*$ heatmaps for scenario 2 for the impulsivity metrics presented in Tables \ref{tab1}}
\label{fig:heatmap2}
\end{figure}

For the first group of statistics described in Section \ref{sec:measures-group1} the results corresponding to Scenario~2 are presented in Fig.~ \ref{fig:heatmap2}. It is worth noting that in this case impulses with amplitudes $A \leq 8$ remain at or below the level of the periodic component present in the reference signal, making them significantly more difficult to detect.

Despite these more challenging conditions, KURT and SKEW still maintain strong discriminative capability. However, their performance is slightly degraded compared to Scenario~1, as reflected by slightly lower values of the aggregated metrics in Table \ref{tab:u_star_summary_scenario2}. Again, energy-based measures (RMS and SD) require a larger number of impulses or higher amplitudes to achieve reliable separation, and the decrease in performence is more visible compared to Scenario~1.

\begin{table}[h!] 
\centering 
\caption{$U^*$-based performance metrics for Scenario~2 for the impulsivity metrics presented in Table \ref{tab1}.}
\label{tab:u_star_summary_scenario2}
\begin{tabular}{l c c c c c}
\hline
Metric & Mean imp. & Mean hard & Mean easy & Coverage & Worst case \\
\hline
KURT & \textbf{0.984} & \textbf{0.965} & \textbf{1.000} & \textbf{0.931} & \textbf{0.565} \\
SKEW & 0.960 & 0.925 & 0.991 & 0.885 & 0.520 \\
RMS & 0.865 & 0.798 & 0.923 & 0.664 & 0.530 \\
SD & 0.864 & 0.797 & 0.923 & 0.664 & 0.530 \\
\hline
\end{tabular}
\end{table}

In Fig.~~\ref{fig:heatmap4x4_scenario1} in in \ref{app:figures}, the corresponding heatmaps for the remaining statistics in Scenario~2 are presented.  The aggregated results are summarized in Table~\ref{tab:u_star_all} in in \ref{app:figures}. Overall, the ranking of statistics remains consistent with Scenario~1, with the exception of the second group of statistics, where SF outperforms CF under the considered conditions. At the same time, a shift in the detection threshold can be observed, as the aggregated performance metrics indicate a slight decrease in sensitivity across all measures. A distinct behavior is observed for CF and ENT, which exhibit a sharp transition in the heatmaps. These metrics remain insensitive when additional impulses are masked by the periodic component but respond very rapidly once the impulses exceed this level, leading to a sudden increase in $U^*$ values even for $n=1$. The decrease in the values of coverage and mean hard for CF and ENT, in comparison to Scenario 1, confirms that their performance is strongly affected by the masking effect of the periodic component, despite maintaining very high value of mean easy. Overall, while the general ranking of statistics remains similar to Scenario 1, the results highlight a reduced sensitivity in the low-amplitude regime and a shift of the effective detection threshold.

\subsection{Impulsivity criteria selection -- results for Scenario 3}
\label{sec:imp_crit_s3}
\begin{figure}[h!]
    \centering
    \includegraphics[width=0.8\textwidth]{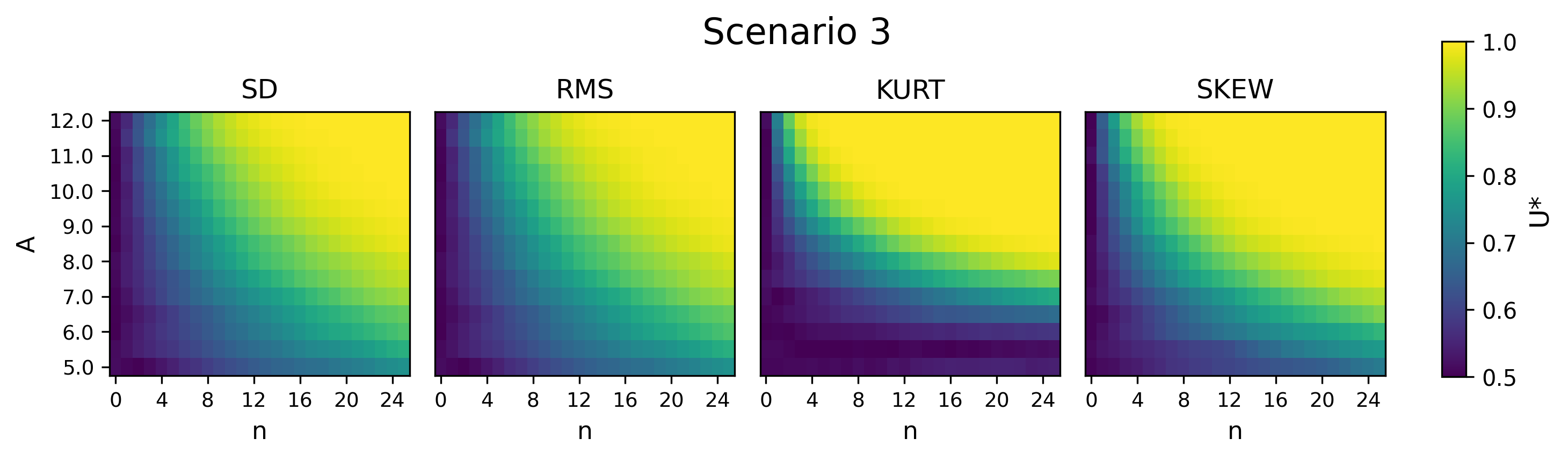}
    \caption{Comparison of $U^*$ heatmaps for Scenario 3 for the impulsivity metrics presented in Tables \ref{tab1}}
\label{fig:heatmap3}
\end{figure}

In Fig.~~\ref{fig:heatmap3} the $U^*$ heatmaps for the impulsivity statistics from Section \ref{sec:measures-group1} in Scenario~3 are presented. In this case, although the nominal amplitude of the periodic component is $I=8$, its modulation results in local amplitude increases up to approximately $9$. Consequently, impulsive components with the values of $A$ below this level are effectively masked by the background signal, making their detection significantly more challenging.

As a result, a further shift of the detection threshold is observed compared to Scenarios~1 and~2, which is clearly reflected in the heatmaps. All metrics require higher impulse amplitudes and/or a larger number of impulses to achieve high $U^*$ values. This leads to a visible reduction in the region of strong separability. Among the analyzed measures from the first group, SKEW achieves the best overall performance, maintaining the highest values of aggregated metrics, see Table \ref{tab:u_star_summary_scenario3}, indicating relatively robust discrimination even under strong masking conditions. SD and RMS exhibit nearly identical behavior, with moderate performance and a smooth transition across the $(A,n)$ space, but reduced sensitivity compared to previous scenarios. 

It is important to notice that KURT, which performed best in Scenarios~1 and~2, shows a noticeable degradation in Scenario~3. Overall, SKEW becomes the best-performing metric within this group, mostly because of its better performance in the hard region. This behavior can be explained by the fundamentally different nature of the background signal. Here, the perodic components produce continuous amplitude variations that can locally exceed the impulse amplitudes for some sets of parameters. Consequently, impulsive events no longer appear as distinct outliers but rather blend into the background, reducing the prominence of heavy tails. In this case, skewness remains sensitive to the resulting asymmetry of the distribution, even when impulsive components are partially masked. The cumulative effect of multiple small or moderate impulses introduces a systematic distortion of the distribution shape, which is more effectively captured by SKEW than by KURT. Consequently, SKEW provides more stable discrimination under conditions where impulsive events do not generate sufficiently strong outliers but still alter the overall distribution.

\begin{table}[h!]
\centering 
\caption{$U^*$-based performance metrics for Scenario~3 for the impulsivity metrics presented in Table \ref{tab1}.}
\label{tab:u_star_summary_scenario3}
\begin{tabular}{l c c c c c}
\hline
Metric & Mean imp. & Mean hard & Mean easy & Coverage & Worst case \\
\hline
KURT & 0.796 & 0.630 & \textbf{0.941} & 0.518 & 0.500 \\
SKEW   & \textbf{0.840} & \textbf{0.734} & 0.933 & \textbf{0.587} & \textbf{0.512} \\
RMS  & 0.808 & 0.728 & 0.878 & 0.521 & 0.503 \\
SD   & 0.808 & 0.728 & 0.878 & 0.521 & 0.503 \\
\hline
\end{tabular}
\end{table}

In Fig.~~\ref{fig:heatmap4x4_scenario1} in in \ref{app:figures}, the corresponding heatmaps for the remaining statistics in Scenario~3 are presented, with the aggregated results summarized in Table~\ref{tab:u_star_all} in in \ref{app:figures}. Compared to Scenarios~1 and~2, a further degradation in performance is observed across all measures. A behavior similar to Scenario~2 is observed for CF and ENT, which again exhibit a sharp transition in the heatmaps. However, this effect is now more pronounced, as these metrics remain insensitive over an even larger region and respond only once the impulse amplitude exceeds the modulated background level. A further insight can be obtained by comparing the best-performing metrics across scenarios. In Scenario~3, the overall ranking of statistics remains mostly consistent, however, the performance gap between the leading measures is reduced. Metrics such as SF, 2-BETA, JD, and SW remain consistently the most effective, demonstrating robust performance under increasing masking conditions. In contrast, previously dominant measures such as CF, ENT, and KURT exhibit a noticeable degradation, particularly in the hard region.

\subsection{Impulsivity assessment within scenarios} \label{sec:impulsivity_assessment}

In this section, the results of the impulsiveness assessment procedure based on bootstrap methodology (described in Section~\ref{sec:methodology}) are presented. Each of the three considered scenarios is analyzed with respect to the measures identified as the most effective within each category (from Section~\ref{sec:measures}). The measure selection procedure was defined as follows: for each group, the measure exhibiting the highest values across the greatest number of performance metrics (Tables~\ref{tab:u_star_summary}--\ref{tab:u_star_summary_scenario3}) was selected. Accordingly, the measures  \{KURT, CF, 2-BETA, JD, SW\} were selected for Scenario 1 and Scenario 2, while \{SKEW, CF, 2-BETA, JD, SW\} were selected for Scenario~3. Importantly, due to the presence of cyclic dependencies in the data in Scenario 2 and Scenario 3, classical bootstrap methods could not be applied. Instead, a block-based bootstrap approach was used, i.e., a methodology adapted to dependent data. A detailed description of the procedure is provided in  \ref{app:bootstrap_app}. 


According to the discussed methodology, first,  we identify whether the given signal can be considered an impulsive one in comparison to the assumed reference scenario. The corresponding figures that illustrate the percentage of Monte Carlo trials in which a statistically significant change in impulsiveness was detected for each parameter configuration and all discussed scenarios can be found in \ref{app:figures_imp_assessment} (Fig.~~\ref{fig:s1_ap}).  The presented results were obtained from 100 Monte Carlo trials for each parameter configuration defined by the number of impulses and their amplitude.

\begin{figure}[h!]
    \centering
\includegraphics[width=\textwidth]{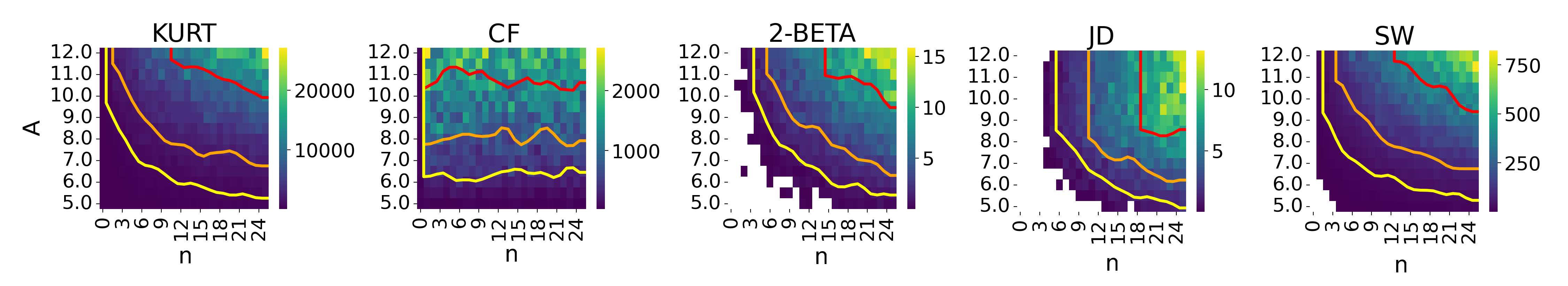}
    \caption{The heatmaps with median of IMF calculated in 100 Monte Carlo trials in Scenario 1. The curves represent smoothing functions and should be interpreted as follows: (a) the red curve corresponds to the 95\% quantile, indicating that 95\% of IMF values lie below it; (b) the orange curve corresponds to the 50\% quantile (median), indicating that 50\% of IMF values lie below it; and (c) the yellow curve corresponds to the 25\% quantile, indicating that 25\% of IMF values lie below it.}
    \label{fig:s1_am}
\end{figure}

\begin{figure}[h!]
    \centering
\includegraphics[width=\textwidth]{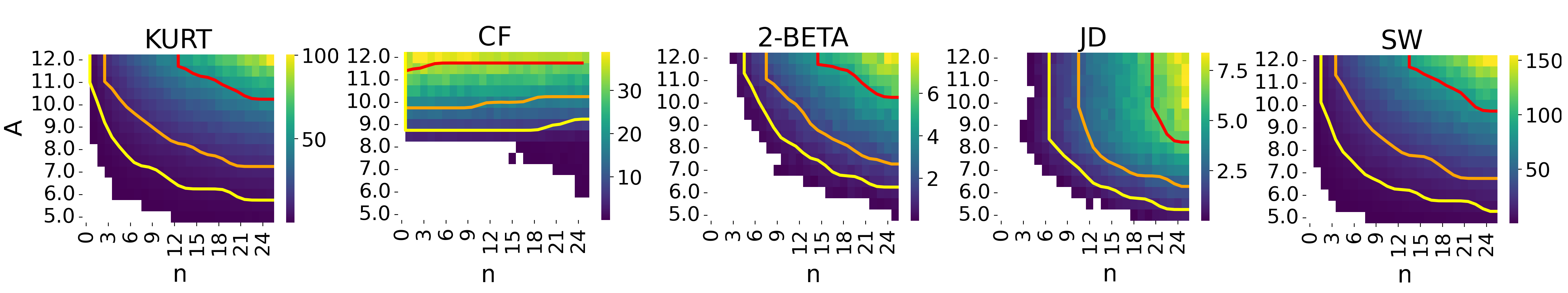}
    \caption{The heatmaps with median of IMF calculated in 100 Monte Carlo trials in the Scanario 2. The curves represent smoothing functions and should be interpreted as follows: (a) the red curve corresponds to the 95\% quantile, indicating that 95\% of IMF values lie below it; (b) the orange curve corresponds to the 50\% quantile (median), indicating that 50\% of IMF values lie below it; and (c) the yellow curve corresponds to the 25\% quantile, indicating that 25\% of IMF values lie below it.}
    \label{fig:s2_am}
\end{figure}

\begin{figure}[h!]
    \centering
\includegraphics[width=\textwidth]{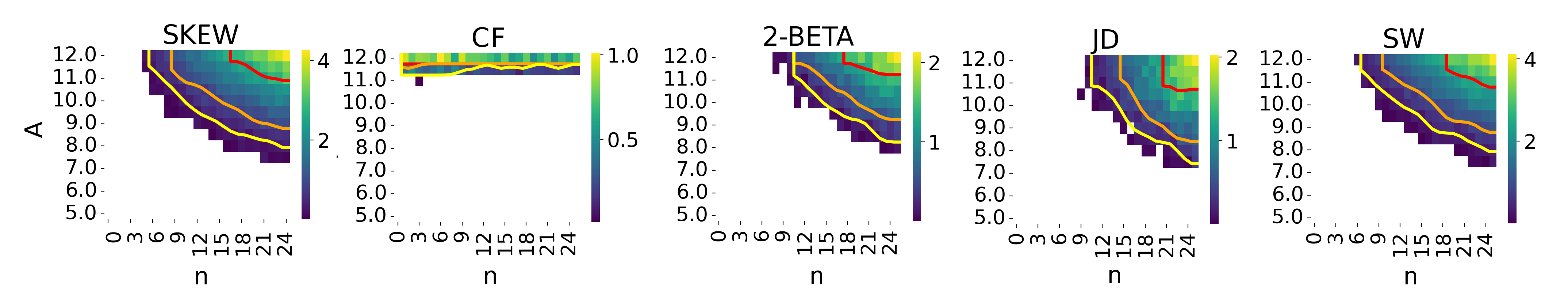}
    \caption{The heatmaps with median of IMF calculated in 100 Monte Carlo trials in Scenario 3. The curves represent smoothing functions and should be interpreted as follows: (a) the red curve corresponds to the 95\% quantile, indicating that 95\% of IMF values lie below it; (b) the orange curve corresponds to the 50\% quantile (median), indicating that 50\% of IMF values lie below it; and (c) the yellow curve corresponds to the 25\% quantile, indicating that 25\% of IMF values lie below it.}
    \label{fig:s3_am}
\end{figure}

The final step of our procedure is the analysis of the IMF behavior to assess the level of detected impulsiveness. Fig.~~\ref{fig:s1_am}--Fig.~~\ref{fig:s3_am} present the median IMF values for selected measures for all scenarios. Similarly, as previously,  the median IMFs were obtained from 100 Monte Carlo trials for each parameter configuration. The parameter grid is consistent with that used in Sections~\ref{sec:imp_crit_s1}--\ref{sec:imp_crit_s3}. White tiles represent cases where no observations were available indicating a change in impulsiveness, i.e., the IMF could not be computed. This outcome is expected when there are no added impulses. The curves represent smoothing functions and should be interpreted as follows: (a) the red curve corresponds to the 95\% quantile, indicating that 95\% of IMF values lie below it; (b) the orange curve corresponds to the 50\% quantile (median), indicating that 50\% of IMF values lie below it; and (c) the yellow curve corresponds to the 25\% quantile, indicating that 25\% of IMF values lie below it\footnote{The quantiles are computed over all simulation cases; if the IMF is undefined for a given case, it is treated as a missing value accounted for in the quantile calculation. Consequently, even if the IMF is defined only for a subset of cases, the quantiles are still evaluated with respect to the full set of simulations.}. 

The median IMF values obtained in Scenario~1 (see Fig.~~\ref{fig:s1_am}) indicate that KURT reflects increasing values of the metric with respect to both the signal amplitude and the number of impulses. The orange curve shows that a median IMF level corresponding to $A=8$ is reached only when the number of impulses $\geq10$. In contrast, for CF, the orange quantile curve remains approximately constant across different values of the number of impulses, corresponding to an amplitude close to $A=8$. This indicates that the IMF behavior for CF differs substantially from that of the other statistics. In this case, the IMF magnitude depends primarily on the amplitude level and is largely insensitive to the density of impulses. Notably, the statistical measures KURT and CF exhibit near-perfect detection of impulsiveness changes when such changes occur (Fig.~~\ref{fig:s1_ap} in \ref{app:figures_imp_assessment}). However, they also produce a high number of false positives, as they indicate significant changes even when the number of impulses equals zero, but the corresponding IMF values remain low. This suggests that these measures are most effective when used as complementary indicators, in conjunction with criteria that first establish the presence of a change. 

More restrictive measures that can effectively serve as indicators include 2-BETA, JD, and SW. As shown in Fig.~~\ref{fig:s1_am}, when the number of impulses equals zero, all these metrics indicate that the IMF is undefined. Importantly, the smallest error is observed for SW; as illustrated in Fig.~~\ref{fig:s1_ap} in \ref{app:figures_imp_assessment}, it allows rejection of the majority of cases with no impulses, while not penalizing (i.e., not rejecting) signals with low impulsiveness (i.e., signals with few impulses with small amplitude). Consequently, SW enables discrimination across a wide range of impulsiveness levels.

In contrast, 2-BETA and JD are more restrictive, tending to classify low impulsiveness as the absence of impulsiveness. However, they allow for clearer differentiation between varying levels of impulsiveness in more pronounced cases. Notably, JD exhibits a slightly different behavior compared to 2-BETA and SW.

As shown in Fig.~~\ref{fig:s1_am}, beyond a certain level, it differentiates impulsiveness more strongly with respect to the frequency of impulses rather than their amplitude. This observation is noteworthy, as distinguishing between different levels of impulsiveness may be of practical importance. Therefore, combining multiple measures can provide more informative insights in impulsiveness assessment, which will be investigated in greater detail in future work.

As shown in Scenario~2, all measures correctly indicate the absence of impulsiveness when the number of impulses equals zero, which is the expected result (see Fig.~~\ref{fig:s1_ap} in \ref{app:figures_imp_assessment}). As illustrated in Fig.~~\ref{fig:s2_am}, the behavior of KURT, 2-BETA, and SW is similar to that observed in Scenario~1, i.e., a gradual increase in IMF values is observed with increasing impulse frequency and amplitude. CF exhibits a comparable trend to Scenario~1, indicating an increase in IMF values with increasing impulse amplitude. However, in this case, a dependence on the number of impulses is also observed - small impulses with low frequency are treated as insignificant, and the IMF is not defined for such cases. An interesting observation arises from the analysis of the percentage of Monte Carlo trials in which a change in impulsiveness was detected for each parameter configuration for IMF values computed using CF (Fig.~~\ref{fig:s1_ap} in \ref{app:figures_imp_assessment}). A clear threshold in detection effectiveness is visible for $A \geq 8$ when the number of impulses is greater than zero. For the remaining statistics, the detection effectiveness exhibits a more gradual behavior: SW and KURT are effective already for a number of impulses greater than zero (although their effectiveness depends on impulse amplitude), whereas 2-BETA and JD become effective for the number of impulses greater than approximately 7, with effectiveness also dependent on impulse amplitude.

As shown in Fig.~~\ref{fig:s3_am}, in Scenario~3 the IMFs computed using CF, 2-BETA, JD, and SW exhibit behavior similar to that observed in Scenario~2. However, the IMF values are determined only for more pronounced impulsive signals, i.e., those characterized by higher impulse frequency and amplitude. In particular, IMFs computed using CF are defined only from approximately $A = 11$, while 2-BETA, JD, and SW exhibit a more gradual response. The IMF computed based on SKEW (the counterpart of KURT in Scenarios~1 and~2) is sensitive to both the frequency and amplitude of impulses.

\section{Real signals analysis} \label{sec:real}
\subsection{Preliminary discussion}
The analyzed datasets were obtained from a condition monitoring system installed on a reciprocating compressor operating in the oil and gas sector. The compressor is designed to increase the pressure of natural gas prior to its transmission through a pipeline to a gas-fired power facility. It is driven by a 2~MW electric motor and consists of four compression cylinders, which progressively elevate the gas pressure in multiple stages. The vibration signals were recorded with a sampling frequency of 25 ~kHz. The dataset includes the vibration measurements acquired from a sensor mounted vertically on one cylinder of the compressor. 

Prior studies on the analysis of simulated data enabled the selection of the most suitable measures for assessing impulsiveness. This was the first step of the general procedure illustrated on the left side of Fig.~ \ref{fig:framework}. As noted, for all the discussed scenarios, the CF, 2-BETA, JD, and SW were selected as the most effective measures from categories 2-5 discussed in Section \ref{sec:measures}. In addition, in Scenario 1 and Scenario 2, the KURT was indicated, whereas in Scenario 3, the SKEW (from category 1).  Thus, finally, for real data, we analyze impulsivity based on the following measures: SKEW, KURT, CF, 2-BETA, JD, SW. 

According to the general methodology proposed, the next step of the analysis is the identification of the impulsivity of the signal and the measurement of the level of detected impulsiveness. This step corresponds to the methodology illustrated on the right hand side of Fig.~ \ref{fig:framework}. The analysis presented in this section corresponds to this step of the procedure and in this regard, it involves identifying impulsivity (0-1 binary information based on $\mathcal{M}_{\mathrm{obs}}$) and assessing its level (by IMF).  

We consider two real data examples from the system; we call them Example 1 and Example 2. In each example, we consider ten 1s signals (nine test signals and one reference signal). Signals analyzed in both examples correspond to reference Scenario 3 discussed in the simulation part, as the signals describe the vibrations of the compressor. In both cases, we analyze $1$s signals from one sensor from different time periods.  However, there is a distinction in the nature (and origin) of the impulsive behaviors relative to the reference. In Example 1, the impulses are symmetric in nature and result from operation of piston inside cylinder, whereas in Example 2, they are highly asymmetric and  were caused by faulty cabling errors. 

\subsection{Example 1}

\begin{figure}[h!]
    \centering
    \includegraphics[width=0.9\textwidth]{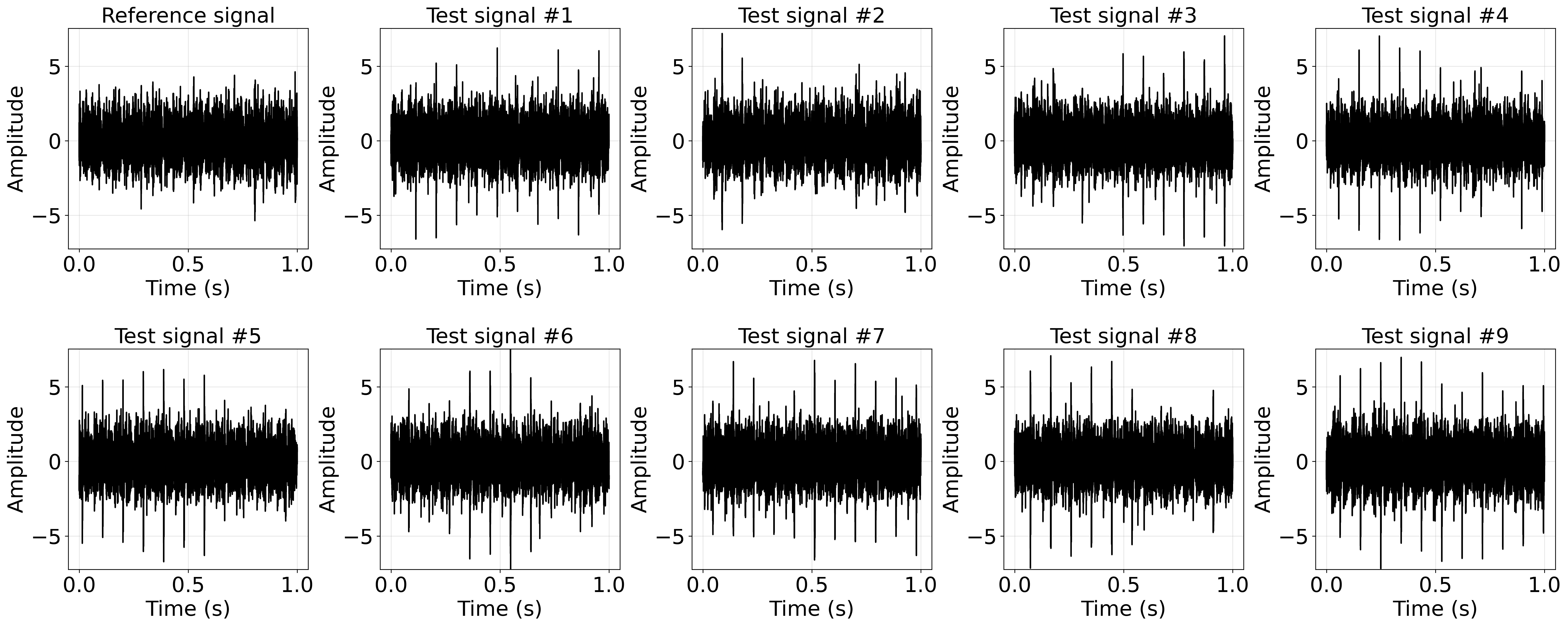}
    \caption{Example 1: Reference signal together with nine test signals.}
    \label{fig:signal_1_signals}
\end{figure}

The analyzed signals are presented in Fig.~\ref{fig:signal_1_signals}.
One of them was selected as the reference signal, while the remaining ones were treated as test signals. As can be noted, the reference signal contains relatively weak impulsive components, whereas the subsequent test signals contain stronger impulsive disturbances, which can be concluded from the visual inspection of the plots. Namely, the number and amplitude of impulsive events increase for signals \#1-\#9 compared to reference signal. In this case, the analyzed impulsiveness is not mainly related to distribution asymmetry, but rather to the occurrence of additional high-amplitude events superimposed on the cyclic background signal.

\begin{figure}[h!]
    \centering
    \includegraphics[width=0.5\textwidth]{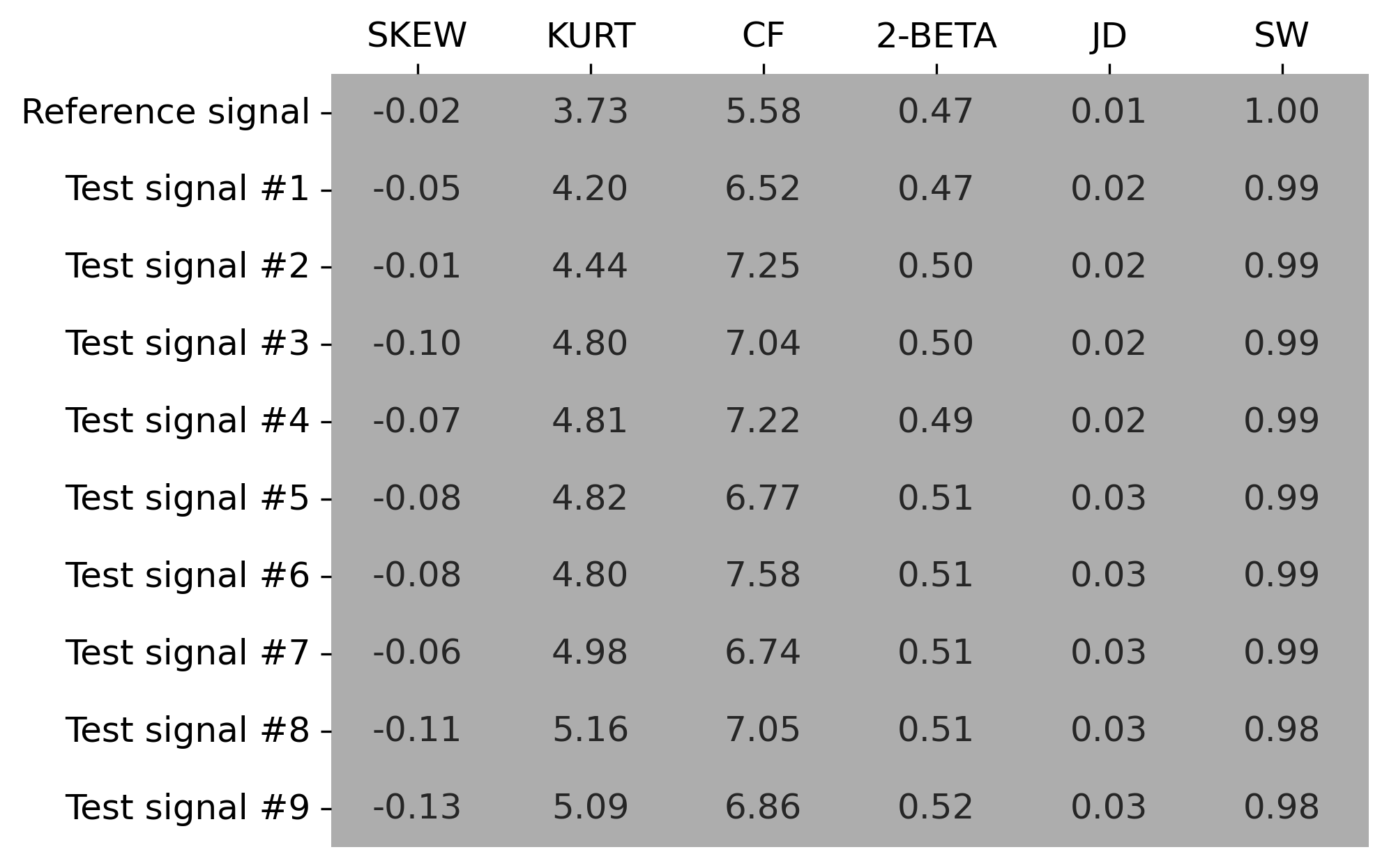}
    \caption{Example 1: Values of selected impulsiveness measures obtained for test signals. 
}
    \label{fig:signal_1_metrics}
\end{figure}

A classical approach to impulsiveness assessment consists in directly comparing the values of such statistics and ordering the signals from the least impulsive to the most impulsive one by using the statistics values. The values of the discussed statistics for all signals are presented in Fig.~\ref{fig:signal_1_metrics}. It is worth to mention that some of the considered measures possess intuitive reference values. For example, skewness values close to zero indicate approximately symmetric amplitude distributions, while kurtosis equal to $3$ corresponds to Gaussian behavior. In practical applications the values of these statistics obtained for the reference signal may differ from such theoretical reference levels due to the specific structure and operating conditions of the analyzed system. In the considered example, this problem is partially solved by the presence of a statistics values for the reference signal to which statistics values of all test signals can be compared. Nevertheless, it is still unclear from which threshold value a signal should be considered impulsive and whether the observed differences between signals are statistically significant.

As can be seen in Fig.~\ref{fig:signal_1_metrics}, the values of SKEW remain relatively close to zero for all analyzed signals. This observation is consistent with the visual inspection of the signals, where impulsive events are approximately symmetric with respect to zero. Consequently, the SKEW measure does not change substantially across the analyzed signals. This behavior is natural, since skewness is expected to be sensitive mainly to asymmetric impulsive disturbances, whereas the impulsive components observed here are predominantly symmetric. A different behavior can be observed for KURT. First of all, for the reference signal, the kurtosis value is already exceeding 3, indicating a deviation from Gaussian behavior caused by the cyclic impulsive structure of the signal.  For consecutive test signals, the kurtosis values increase compared to the reference signal, suggesting progressive changes in the tail thickness of the corresponding amplitude distributions. However, based solely on the raw values of KURT, it is difficult to determine whether the observed differences are significant or simply result from the natural variability of the signal. Similar observations can be made for CF. The obtained values increase for the test signals, which is expected since CF is strongly influenced by larger signal amplitudes. This might suggest the presence of stronger impulsive events in the test signals. However, similarly to KURT, the interpretation of the observed differences remains ambiguous without statistical assessment. For 2-BETA, JD, and SW, the interpretation of the raw statistic values is even less straightforward. Although these measures were identified in the simulation study as highly effective for distinguishing impulsive and non-impulsive signals, their numerical values for the analyzed real signals remain relatively close to one another. Consequently, the impulsiveness level cannot be reliably assessed directly from their values alone.

\begin{figure}[h!]
    \centering

    \begin{subfigure}[t]{0.5\textwidth}
        \centering
        \includegraphics[width=\linewidth]{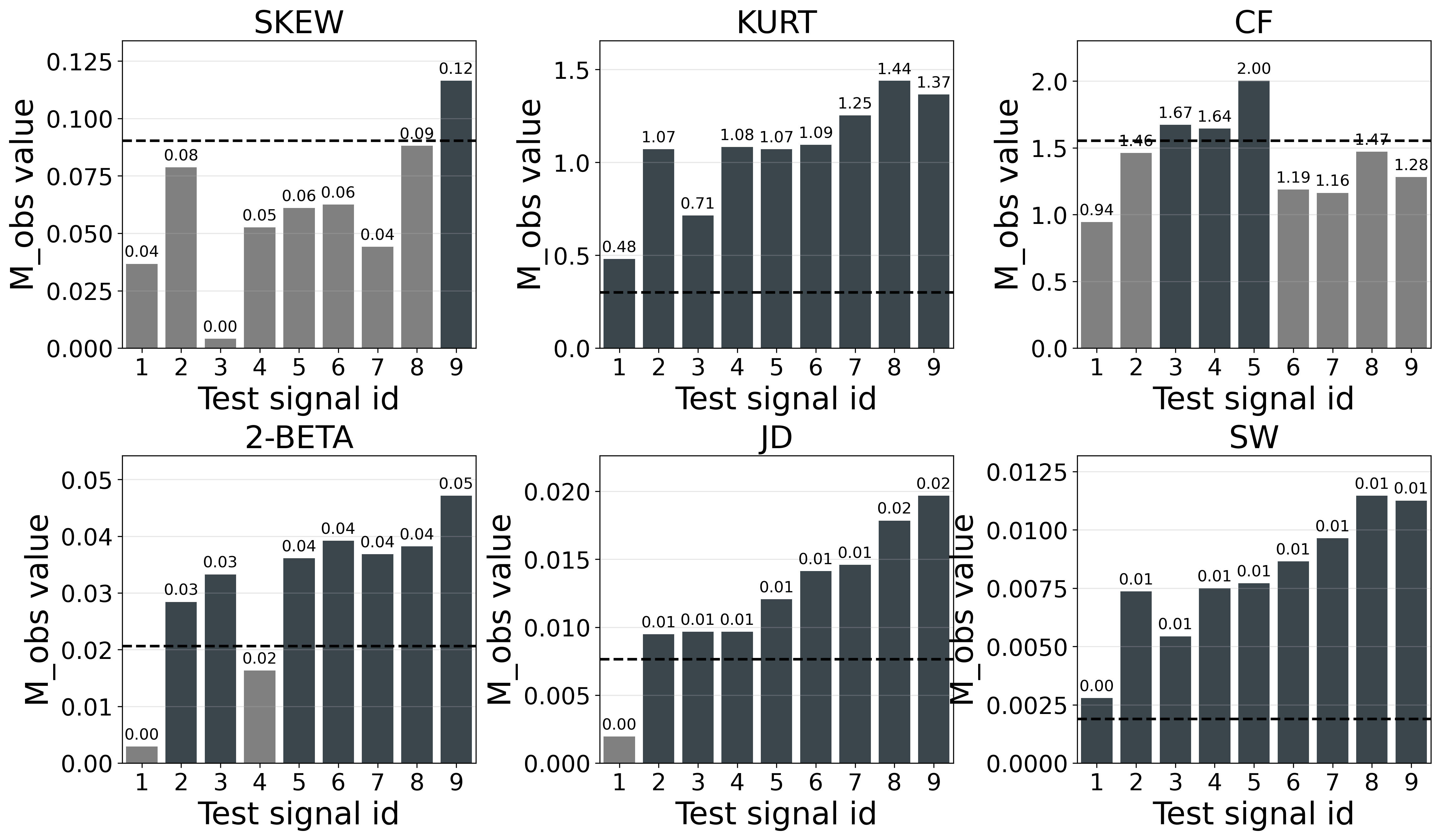}
        \caption{$M_{obs}$ values obtained in the bootstrap procedure for individual test signals and selected impulsiveness measures. The dashed line indicates $M_c$ determined from the reference signal. Gray bars denote statistically insignificant $M_{obs}$ values, whereas black bars denote statistically significant values.}
        \label{fig:signal_1_barplots_a}
    \end{subfigure}
    \hfill
    \begin{subfigure}[t]{0.49\textwidth}
        \centering
        \includegraphics[width=\linewidth]{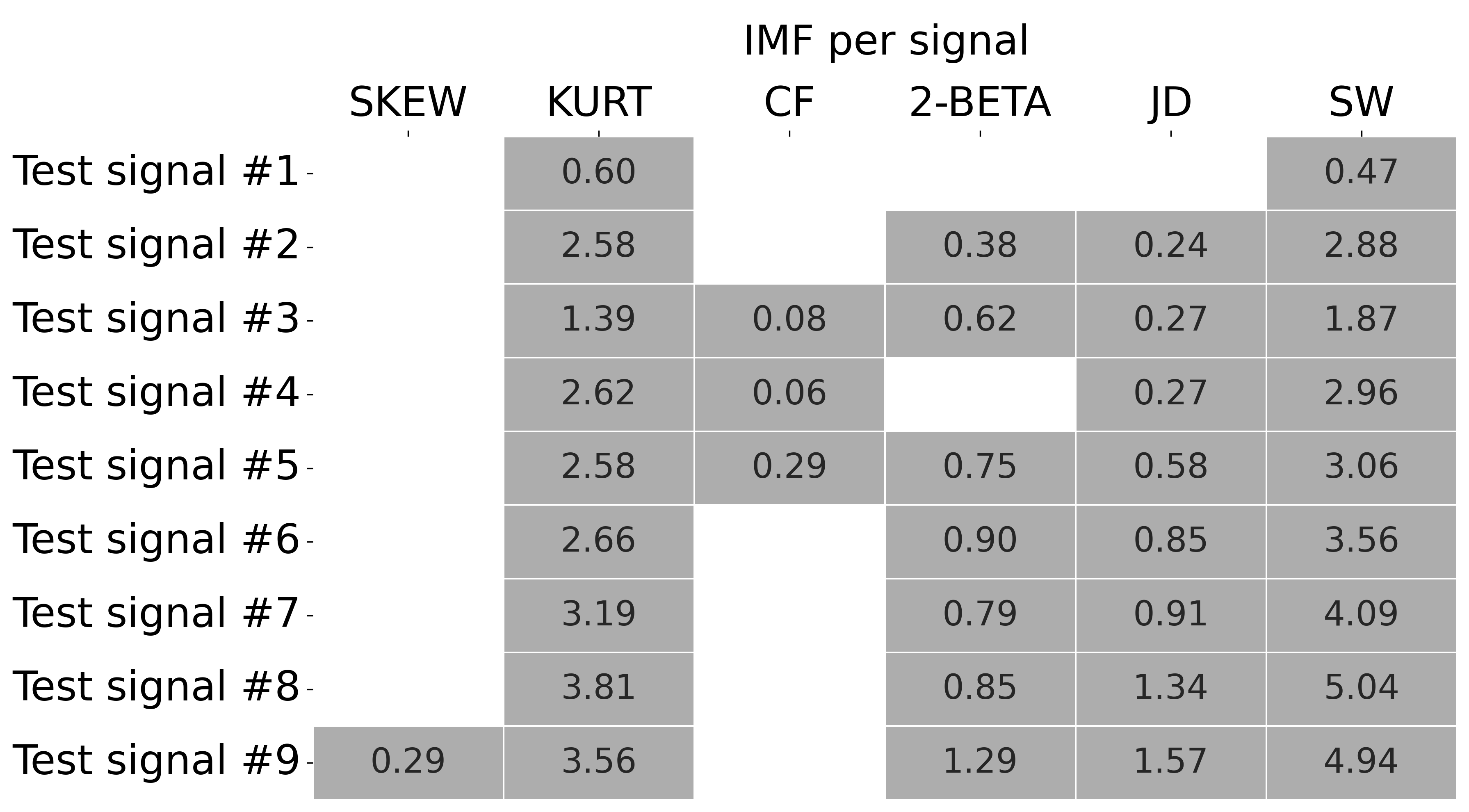}
        \caption{IMF values corresponding to individual impulsiveness measures and test signals. White cells indicate that no statistically significant change was detected in the signal and therefore IMF was not determined.}
        \label{fig:signal_1_barplots_b}
    \end{subfigure}

    \caption{Results for Example 1. Detailed descriptions of panels (a) and (b) are provided in their respective captions.}
    \label{fig:signal_1_barplots}
\end{figure}

In the next step of the analysis, we apply the BIA procedure introduced in Section \ref{BIAA}. The goal of this analysis is to determine whether the differences between the reference signal and the test signals are significant with respect to the considered impulsiveness measures and to assess the strength of the detected impulsiveness changes. The obtained $\mathcal{M}_{\mathrm{obs}}$ values determined using the BIA procedure are presented in Fig.~\ref{fig:signal_1_barplots_a}. The dashed line corresponds to the critical value $\mathcal{M}_c$ (considered here as the threshold) obtained from the reference signal using the bootstrap procedure. Values exceeding this threshold indicate statistically significant differences between the reference and the analyzed test signal with respect to the considered impulsiveness measure.

First of all, the results demonstrate that the statistical significance assessment substantially changes the interpretation of the analyzed measures. In particular, for 2-BETA, JD, and SW, the raw statistic values, shown previously in Fig.\ref{fig:signal_1_metrics}, appeared relatively close to one another, making direct interpretation difficult. However, the BIA analysis reveals that even these nominally small differences are  significant (statistically). Consequently, the analyzed signals are distinguishable with respect to the characteristics measured by these statistics, i.e., differences in the overall distribution structure, deviations from the reference distribution, and changes related to impulsive heavy-tailed behavior. At the same time, the relatively small variations observed for SKEW are confirmed to be statistically insignificant for most of the analyzed signals. This observation is consistent with the approximately symmetric nature of the impulsive disturbances, which only weakly affect the asymmetry of the amplitude distribution. 

The next step of the analysis involves assessing the level of the detected impulsiveness changes. To this end, the IMF index is calculated for all statistically significant cases. We recall, the IMF represents the relative increase of impulsiveness with respect to the reference signal and is determined separately for each impulsiveness measure. In contrast to the raw statistic values, which only describe the absolute value of a given impulsiveness measure, the IMF provides a normalized description of the impulsiveness increase relative to the reference signal. In simple terms, an IMF value equal to $x$ means the impulsive characteristic is approximately $x$ times stronger than in the reference signal. However, this does not mean the raw impulsiveness measure is $x$ times larger; instead, the IMF quantifies the relative strength of the change compared to the reference signal's natural variability. Consequently, IMF enables ordering the analyzed signals from the least impulsive to the most impulsive ones in the sense of a given impulsiveness measure, while simultaneously taking into account the statistical significance and relative strength of the observed impulsiveness changes. The obtained IMF values are presented in Fig.~\ref{fig:signal_1_barplots_b}. Blank white cells indicate cases for which no statistically significant impulsiveness change was detected, and therefore IMF was not calculated.  It can be observed that the ordering of the signals (with respect to the impulsiveness level) depends on the selected impulsiveness measure. 

The IMF-based analysis is consistent with the conclusions drawn previously from the statistical significance analysis. In the case of SKEW, almost all analyzed signals are classified as non-impulsive with respect to the reference signal, except for signal \#9, which is classified as possessing a low impulsiveness level. This observation is consistent with the approximately symmetric nature of the impulsive disturbances, which only weakly affect the asymmetry of the amplitude distribution. For KURT, all signals are classified as statistically significant signals, most of them are highly impulsive. Moreover, the IMF values indicate that the strongest increase of tail-related impulsiveness occurs for signals \#7--\#9. The same conclusions can be drawn from how SW classifies signals. A different behavior can be observed for CF. In this case, only signals \#3-\#5 are classified as impulsive, with lower level of impulsivness. This observation confirms that CF primarily emphasizes sparse isolated extreme amplitudes rather than dense impulsive behavior. For 2-BETA, most of the statistically significant signals are classified as possessing low impulsiveness levels. This suggests that this measure is effective for detecting statistically significant deviations from the reference distribution, but provides limited discrimination capability with respect to the impulsiveness level itself. Similar conclusions we have for JB-based IMF values. 

As a summary of this example, we identify the signals $\#2-\#9$ as impulsive (based on KURT, 2-BETA, JD, and SW). Additionally, KURT and SW also indicate signal $\#1$ as impulsive (however, with a low impulsivity level). The analysis presented here reveals an interesting observation: two signals with relatively similar raw statistical values can yield substantially different IMF values if the changes differ in statistical significance or relative deviation from the reference case.

\subsection{Example 2}

\begin{figure}[h!]
    \centering
    \includegraphics[width=0.9\textwidth]{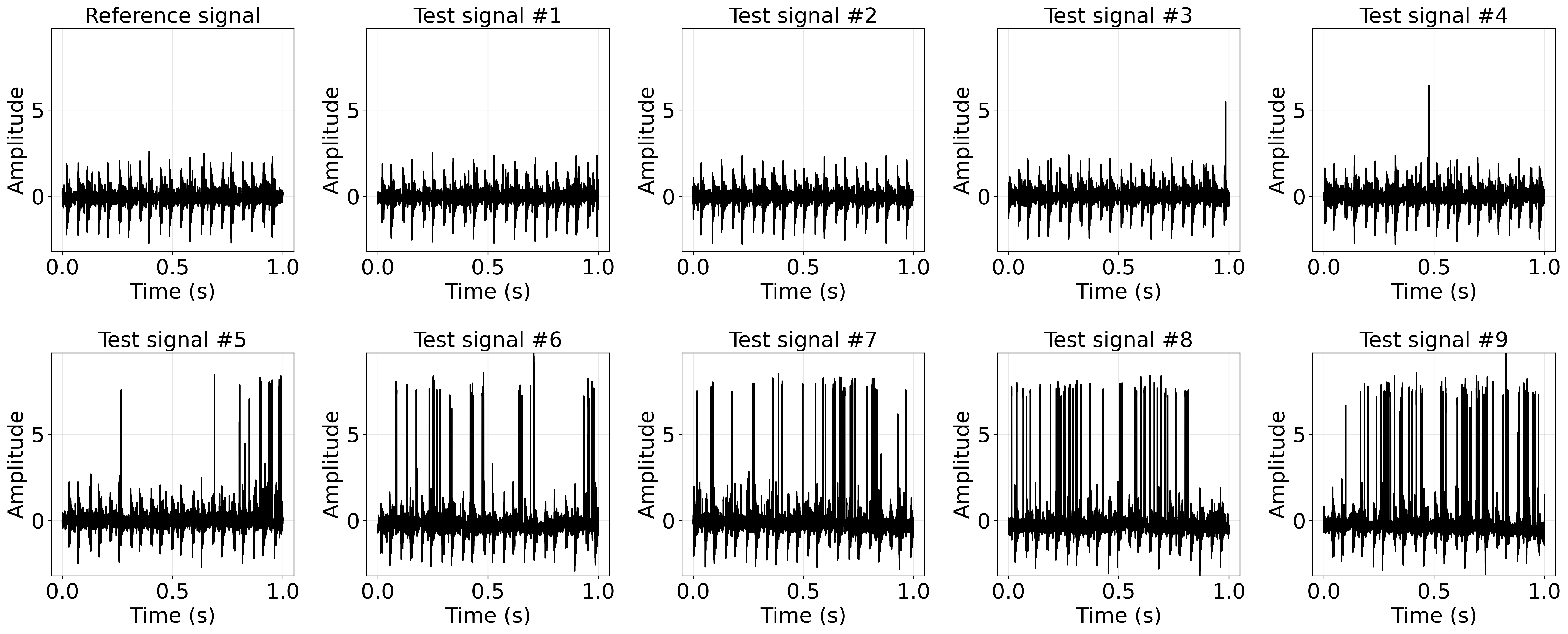}
    \caption{Example 2: Reference signal together with nine test signals.}
    \label{fig:signal_2_signals}
\end{figure}

The analyzed signals corresponding to this example are presented in Fig. \ref{fig:signal_2_signals}. Similarly to Example 1, one signal was selected as the reference signal, while the remaining ones were treated as test signals. In this case, however, the impulsive behavior differs substantially from the previous example. Here, the impulsive disturbances are strongly asymmetric and dominated by high positive-amplitude impulses. Moreover, starting from signal \#5, the number and density of impulsive events increase significantly, which is clearly visible during visual inspection of the signals.

\begin{figure}[h!]
    \centering
    \includegraphics[width=0.5\textwidth]{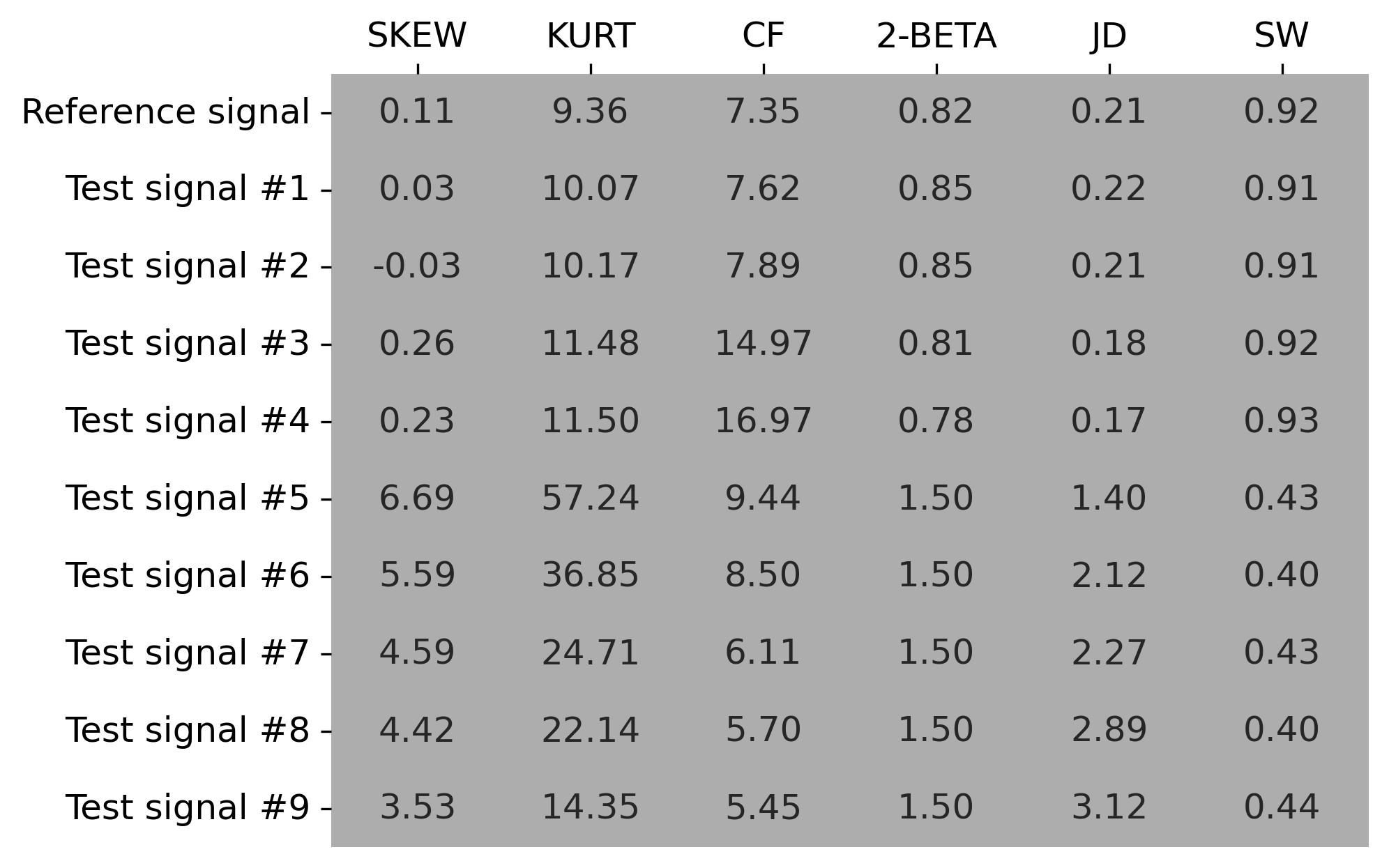}
    \caption{Example 2: Values of selected impulsiveness measures obtained for test signals.
}
    \label{fig:signal_2_metrics}
\end{figure}
The values of the selected measures calculated for all analyzed signals are presented in Fig. \ref{fig:signal_2_metrics}. Contrary to Example 1, several measures already suggest substantial impulsive changes directly from their raw values. In particular, SKEW, 2-BETA, JD and SW indicate a pronounced transition between signals \#1--\#4 and signals \#5--\#9. This behavior is consistent with the visual appearance of the signals, where the latter group contains dense high-amplitude impulsive disturbances. 

\begin{figure}[h!]
    \centering

    \begin{subfigure}[t]{0.5\textwidth}
        \centering
        \includegraphics[width=\linewidth]{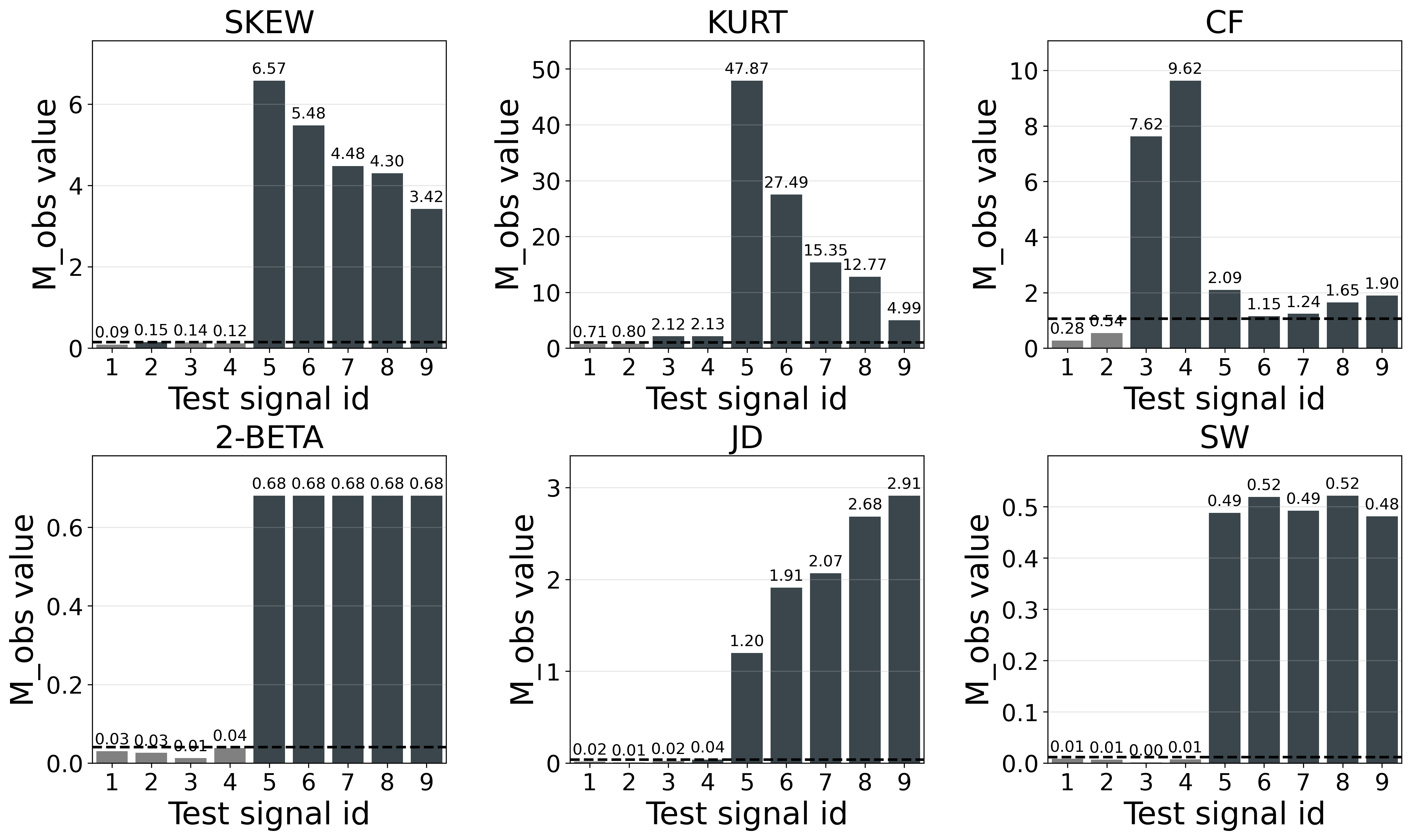}
        \caption{$M_{obs}$ values obtained in the bootstrap procedure for individual test signals and selected impulsiveness measures. The dashed line indicates $M_c$ determined from the reference signal. Gray bars denote statistically insignificant $M_{obs}$ values, whereas black bars denote statistically significant values.}
        \label{fig:signal_2_barplots_a}
    \end{subfigure}
    \hfill
    \begin{subfigure}[t]{0.48\textwidth}
        \centering
        \includegraphics[width=\linewidth]{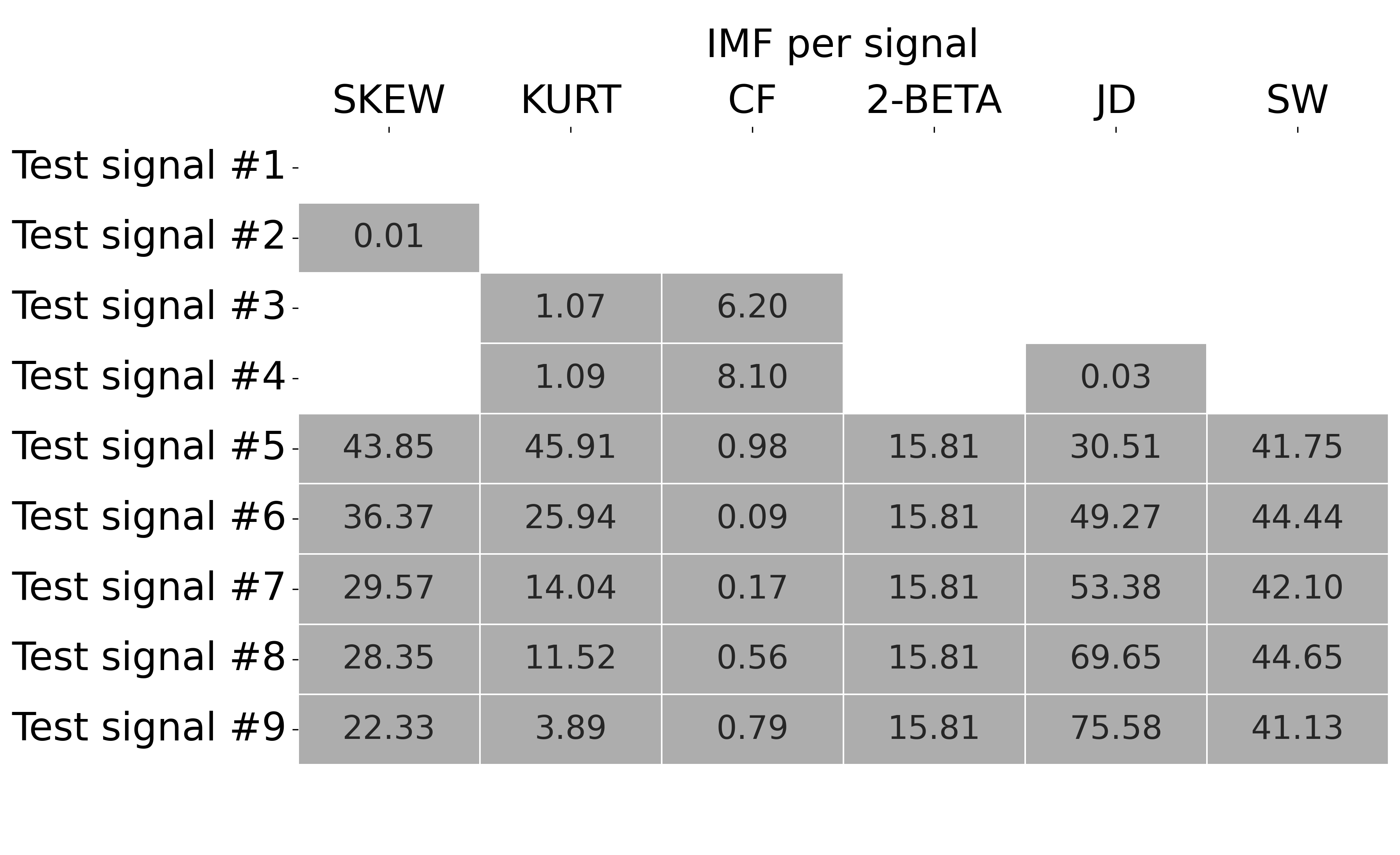}
        \caption{IMF values corresponding to individual impulsiveness measures and test signals. White cells indicate that no statistically significant change was detected in the signal and therefore IMF was not determined.}
        \label{fig:signal_2_barplots_b}
    \end{subfigure}

    \caption{Results for Example 2. Detailed descriptions of panels (a) and (b) are provided in their respective captions.}
    \label{fig:signal_1_barplots}
\end{figure}

According to the proposed methodology, the next step consists in statistical assessment of the observed impulsiveness changes using the $\mathcal{M}_{obs}$ statistics obtained within the BIA framework. The corresponding results are shown in Fig.~\ref{fig:signal_2_barplots_a}. We note, they confirm that signals starting from \#5 are characterized by statistically significant impulsiveness changes for all considered  measures.

One can also see that at the same time, the individual statistics react differently to the analyzed impulsive behavior. The SKEW-based $\mathcal{M}_{obs}$ values increase strongly for signals \#5--\#9, which is expected since the impulsive disturbances are strongly asymmetric. However, the corresponding values gradually decrease for the last signals despite the increasing number of impulses visible in Fig.~\ref{fig:signal_2_signals}. This effect results from the fact that skewness is primarily sensitive to asymmetry caused by sparse extreme observations. As the impulse density increases, the impulsive component becomes less isolated and contributes more strongly to the overall distribution variance, which reduces the relative asymmetry effect captured by SKEW.

A similar phenomenon can be observed for KURT. The largest $\mathcal{M}_{obs}$ values are obtained for signals \#5 and \#6, whereas for signals \#7--\#9 the corresponding values decrease despite visually stronger impulsive behavior. This observation is related to the interpretation of kurtosis as a measure of heavy-tailedness caused mainly by rare extreme observations. For sparse impulsive signals, individual impulses strongly dominate the distribution tails, leading to very high kurtosis values. However, when impulses become dense, large amplitudes are no longer rare events and therefore the relative heaviness of the tails decreases from the statistical point of view. Consequently, signals containing many impulses may visually appear more impulsive while simultaneously producing lower kurtosis-related impulsiveness indices. The CF-based results exhibit yet another type of behavior, namely, for signals \#3 and \#4, where isolated high-amplitude impulses are clearly visible. Since CF depends on the ratio between the maximal amplitude and the RMS value, sparse extreme peaks produce the strongest response. In contrast, for signals \#5--\#9, although the impulse amplitudes remain high, the increasing number of impulses also increases the RMS level, which reduces the relative contribution of individual peaks and therefore lowers the CF-based impulsiveness indication. The behavior of JD and SW is considerably more consistent with the visual inspection of the signals. Both measures indicate increasing statistically significant deviations from the reference distribution starting from signal \#5, and the corresponding $\mathcal{M}_{obs}$ values gradually increase for consecutive signals. This observation confirms that these measures are sensitive not only to isolated extreme observations but also to more global changes in the overall distribution structure caused by dense impulsive disturbances. The 2-BETA statistic also successfully identifies all highly impulsive signals. However, its discrimination capability remains limited because the obtained $\mathcal{M}_{obs}$ values are numerically very similar for all highly impulsive cases.
The corresponding IMF values are presented in Fig.~\ref{fig:signal_2_barplots_b}. Similarly to the previous example, empty cells correspond to statistically insignificant cases for which IMF was not calculated. The obtained results confirm the conclusions drawn from the $\mathcal{M}_{obs}$ analysis while additionally providing information regarding the relative level of impulsiveness changes with respect to the reference signal. The most intuitive and visually consistent results are obtained for JD and SW. In both cases, signals \#5--\#9 are classified as moderate or highly impulsive, and the corresponding IMF values increase progressively together with the visible increase in impulsive disturbance density. Importantly, this behavior is not directly visible when analyzing only the raw values of these statistics shown in Fig.~\ref{fig:signal_2_metrics}. The impulsiveness ordering becomes clear after introducing the statistical significance assessment and the IMF-based normalization. For SKEW and KURT, the IMF-based ordering again differs from the visual impulsiveness ordering observed in Fig.~\ref{fig:signal_2_signals}. In particular, the largest IMF values are obtained for signals containing more sparse asymmetric impulses rather than for signals characterized by the largest impulse density.


As a summary of this example, the signals \#5--\#9 can be identified as impulsive with respect to all considered impulsiveness measures. Additionally, KURT and CF also classify signals \#3 and \#4 as impulsive. This behavior is consistent with the statistical interpretation of these measures, since both KURT and CF are particularly sensitive to sparse isolated high-amplitude impulses. Consequently, these measures indicate impulsiveness already for signals containing only several strong impulsive events. At the same time, however, they do not necessarily provide the intuitive impulsiveness gradation for signals containing dense impulsive disturbances - the corresponding IMF values decrease despite the visibly increasing number of impulsive events - and although they are highly sensitive detectors of isolated impulsive events, they may be less suitable for assessing impulsiveness growth associated with increasing impulse density. The obtained results demonstrate therefore that different impulsiveness measures may emphasize different statistical aspects of impulsive behavior and may lead to partially different impulsiveness classifications. As a consequence, reliable impulsiveness assessment should not be based on a single statistic alone. Instead, meaningful interpretation of impulsive behavior often requires joint analysis of several complementary impulsiveness measures. Our analyzes, based on the BIA methodology, confirm the preliminary conclusions obtained from the raw statistical values alone; therefore, in this case, they can be viewed as complementary to the classical approach.

\section{Summary and conclusions}
This article addresses the critical challenge of identifying and quantifying impulsivity in diagnostic signals and its change in comparison to the reference signal, particularly within the context of condition monitoring. While impulsive behavior is often inherent in industrial processes, such as the operation of reciprocating compressors, it can also stem from external disturbances or measurement errors. To evaluate the change in impulsiveness of given signal is vital, as the presence of high-intensity impulsivity often renders classical Gaussian-based analytical methods ineffective and requires the use of more computationally intensive robust techniques.

To solve this, we introduced a statistically grounded, two-stage framework:
\begin{itemize}
\item Metric selection: we systematized a wide array of impulsivity measures into five categories (basic statistics, combinations, distribution parameters, distance measures, and test statistics). Using an objective selection criterion based on the Mann-Whitney statistic, we identified the "best-performing" measures for specific signal environments, effectively removing the subjectivity inherent in choosing a metric.
\item Statistical quantification: we established a formal procedure using bootstrap-driven resampling to determine the statistical significance of detected impulsivity. Furthermore, we defined an impulsiveness magnitude factor to provide a relative measure of intensity, allowing for a standardized comparison across different metrics.
\end{itemize}

The proposed framework offers a scalable and objective tool for signal characterization. By providing both a binary identification of impulsivity (based on BIA methodology) and a quantified magnitude index (IMF), this research enables practitioners to make informed decisions regarding the necessity of advanced robust methods, thereby optimizing the reliability and efficiency of diagnostic workflows.


The methodology was validated through extensive Monte Carlo simulations across three reference scenarios, ranging from a baseline stochastic model to a complex modulated signal. The simulation study demonstrates that the proposed framework effectively distinguishes between different signal scenarios, identifying the most robust impulsivity metrics across varying background conditions. In Scenario 1 and Scenario 2, kurtosis and crest factor exhibit superior discriminative power, though CF remains primarily sensitive to impulse amplitude rather than density. In the more complex Scenario 3, where periodic disturbances generate oscillatory bursts that may mask additional disturbances, skewness outperforms kurtosis by effectively capturing the resulting distribution asymmetry. Overall, the bootstrap-based magnitude analysis reveals that while metrics like KURT and CF are highly sensitive, they can produce false positives; consequently, more restrictive measures such as 2-BETA, JD, and the SW statistic are identified as more reliable indicators for formally establishing the presence and magnitude of significant impulsivity. The introduced IMF provides a normalized index to quantify the intensity of detected impulsivity relative to the reference signal's natural variability. For Scenarios 1 and 2, the IMF for CF depends primarily on the impulse amplitude and remains largely insensitive to impulse density. In contrast, the IMF for KURT reflects increases in both amplitude and the number of impulses. In a more complex scenario (Scenario 3), the ranking of metrics changes. While IMF for  CF and KURT exhibit performance degradation, the IMF for metrics such as 2-BETA, JD, and SW remain consistently effective at quantifying the magnitude of impulsivity even under increased masking conditions. 

In addition, we provided a discussion of two real-data examples. The presented here analyzes demonstrate that impulsiveness should not be interpreted as a single, one-dimensional property of the signal. Different impulsiveness measures emphasize different statistical characteristics of impulsive behavior and therefore may lead to different impulsiveness rankings for the same set of signals. In particular, as we demonstrated based on real signals analysis,  CF mainly emphasizes isolated extreme amplitudes, KURT is strongly related to heavy-tailed behavior and rare outliers, while SKEW primarily reflects distribution asymmetry. In contrast, JD and SW quantify more global deviations of the signal amplitude distribution from the reference behavior, whereas 2-BETA reflects changes in the tail decay and sharpness of the distribution associated with the generalized Gaussian model. Consequently, signals that are highly impulsive in the sense of one statistical property are not necessarily highly impulsive with respect to another one. 
 In other words, we have shown that the problem of impulsiveness assessment is inherently multi-dimensional. 

Therefore, the goal of the proposed methodology based on statistical-based approach  is not to reduce impulsiveness assessment to a single scalar quantity, but rather to provide a statistically justified framework for identifying and quantifying different types of impulsive behavior. 
In this sense, the proposed approach enables not only the statistical verification of whether the analyzed signals differ from the reference signal, but also the interpretation of which impulsive characteristics are responsible for these differences and how strong these differences are with respect to a selected impulsiveness measure.

The obtained results for real signals clearly show that direct comparison of impulsiveness measure values may lead to incomplete or even misleading conclusions. Different measures operate on different numerical scales and describe different signal characteristics; therefore, the nominal magnitude of the observed differences alone may not provide sufficient information about their significance, as it was demonstrated in real-data Example 1. The performed statistical assessment allows one to determine whether the observed differences actually exceed the natural variability of the measure for the reference signal. {The results for real-data examples based on  the IMF-based analysis may reveal meaningful differences between signals even when the corresponding raw impulsiveness statistics remain numerically similar. At the same time, the results showed that combining multiple complementary measures provides substantially more informative impulsiveness characterization than relying on a single statistic alone, especially for complex industrial signals containing both cyclic and impulsive components.}


Important justification for this research work was its potential for applications in condition monitoring. This becomes especially useful in health assessment of machines, for which impacts are generated in perfectly normal conditions (e.g. reciprocating machinery/ compressor, mining, processing, manufacturing, etc.). Another phenomenon are random impacts, which can be a disturbance and should be neglected or represent an incipient fault. In any case, we need a "measure/ detector" of impulsiveness and such a detection method should be able to operate when the normal healthy signal is of an impulsive nature.  The framework proposed in the paper exhibits such features. The examples of real signals confirmed engineering intuition about impulsiveness analysis being a multi-faceted problem, we now obtained first evidence-based results, which covered real life cases of a mechanical and electronic fault. In the future research, the proposed framework should be further developed to create a flowchart based method, which should be able to:
validate,
discriminate,
and assess level of impulsiveness using the best performed measures. 

Although this study focuses on signals from healthy machines, the proposed methodology could be applicable to signals containing a mixture of impulsive signal of interest and Gaussian noise. This is particularly relevant when a fault emerges in a machine that already generates impulses during normal operation, necessitating the differentiation of multiple signal components. Consequently, proposed approach  can be adapted for fault detection in Gaussian environments. Our current research is going in this direction.

Future work will also investigate the integration of multiple measures-based analysis to provide deeper insights into complex impulsive signal structures. Our idea is also related to the incorporation of intelligent methods for the classification of signals with respect to their impulsivity levels, based on the BIA and IMF-based analyzes. 

\section*{Acknowledgements}
The work of JW, DK, RZ and AW is supported by National Center of Science under Weave-Unisono project No. 2025/07/Y/ST8/00070 ”Advanced signal processing techniques for cyclostationary modelling in Gaussian and non-Gaussian noisy environment -detection of cyclic sources, estimation, optimisation of algorithms and validation in the context of fault identification”.
\bibliography{bibfile2}
\newpage\clearpage\appendix
\section{Definitions of the discussed impulsivity measures} \label{app:tables}

In Tables \ref{tab1}--\ref{tab5}, the following notation is used. 
Let $x_1,\ldots,x_k$ denote the observed signal that is a realization of a random process $X$. The sample mean is defined as
\begin{equation}
\bar{x}=\frac{1}{k}\sum_{i=1}^{k}x_i .
\end{equation}
The symbols $\hat{f}(x)$ and $\hat{F}(x)$ denote the estimated probability density function (PDF) and cumulative distribution function (CDF) of the observed signal, respectively. The functions $f_0(x)$ and $F_0(x)$ denote the corresponding reference probability density and cumulative distribution functions representing an assumed background model (e.g., Gaussian), depending on the application.

\begin{longtable}{|c|m{3cm}|c|>{\centering\arraybackslash}m{3.5cm}|m{7cm}|}
\caption{Basic statistics}
\renewcommand{\arraystretch}{0.8}
\label{tab1}\\
\hline
Id & Name & Abb & Formula / description & Interpretation \\
\hline
\endfirsthead
\hline
Id & Name & Abb & Formula / description & Interpretation \\
\hline
\endhead
\hline
\endfoot
\hline
\endlastfoot
1 & Standard deviation \cite{basic_stats2} & SD &
$\sqrt{\frac{1}{k-1}\sum\limits_{i=1}^{k} (x_i-\bar x)^2}$ &
Quantifies the amount of variation or dispersion in a signal. \\ \hline
2 & Root mean square \cite{basic_stats2} & RMS &
$\sqrt{\frac{1}{k}\sum\limits_{i=1}^{k} x_i^2}$ &
It measures the total  energy of the signal. \\ \hline
3 & Kurtosis \cite{basic_stats2} & KURT &
$\frac{1}{k}\sum\limits_{i=1}^{k}\left(\frac{x_i-\bar x}{SD}\right)^4$ &
It tracks how often rare, extreme values (outliers) occur compared to a Gaussian PDF. \\ \hline
4 & Skewness \cite{basic_stats2} & SKEW &
$\frac{1}{k}\sum\limits_{i=1}^{k}\left(\frac{x_i-\bar x}{SD}\right)^3$ &
Describes the asymmetry of the signal distribution around the mean.\\ \hline
\end{longtable}

\begin{longtable}{|c|m{3cm}|>{\centering\arraybackslash}m{1cm}|>{\centering\arraybackslash}m{4cm}|m{6.5cm}|}
\caption{Combinations of basic statistics}
\renewcommand{\arraystretch}{0.8}
\label{tab2}\\
\hline
Id & Name & Abb & Formula / description & Interpretation \\
\hline
\endfirsthead
\hline
Id & Name & Abb & Formula / description & Interpretation \\
\hline
\endhead
\hline
\endfoot
\hline
\endlastfoot
1 & Shape factor \cite{matlab_documentation} & SF &
$ \frac{\mathrm{RMS}}{\frac{1}{k}\sum\limits_{i=1}^{k}|x_i|}$ &
It measures how much an overall profile of the signal deviates from a perfect, constant flat line. \\ 
\hline
2 & Crest factor \cite{matlab_documentation} & CF &
$\frac{\max |x_i|}{\mathrm{RMS}}$ &
It measures the ratio between the highest instantaneous peak and the overall average intensity (RMS) of a signal. \\ \hline
3 & Entropy \cite{entropy} and Negentropy \cite{negentropy} & ENT and NEG &
$-\sum\limits_{i=1}^{k} \hat{f}(x_i)\log \hat{f}(x_i)$. Negentropy is a difference between ENT calculated for Gaussian distributed sample and ENT calculated for given signal &
ENT is a measure of disorder, randomness, or uncertainty within a given system. NEG is opposite to ENT. \\ \hline
4 & Gini index \cite{gini_index} & GINI &
$\frac{\sum\limits_{i=1}^{k}\sum\limits_{j=1}^{k}|x_i-x_j|}{2k\sum\limits_{i=1}^{k} |x_i|}$ & It measures the level of inequality within a distribution, where a value of 0 represents perfect equality and a value of 1- absolute concentration or inequality. \\ \hline





\end{longtable}

\begin{longtable}{|c|m{3cm}|c|>{\centering\arraybackslash}m{4cm}|m{5.5cm}|}
\caption{Estimated parameters}
\renewcommand{\arraystretch}{0.8}
\label{tab3}\\
\hline
Id & Name & Abb & Formula / description & Interpretation \\
\hline
\endfirsthead
\hline
Id & Name & Abb & Formula / description & Interpretation \\
\hline
\endhead
\hline
\endfoot
\hline
\endlastfoot
1 & $2-\alpha$, where $\alpha$ is the stability index of the  $\alpha-$stable distribution \cite{cvb,mcculloch1986simple} & 2-ALPHA &
Estimated by regression method \cite{kont} &The $\alpha$ parameter ranging from $0$ to $2$, determines how "heavy" is the distribution tail. \\ \hline
2 & $\sigma$, where $\sigma$ is the scale parameter of the  $\alpha-$stable distribution  & SIGMA &Estimated by regression method \cite{kont}&
It quantifies the intensity and spread of the shocks in a signal, serving as a robust alternative to standard deviation for heavy-tailed data \\ \hline
3 & number of degrees of freedom from Student's t degrees distribution \cite{li2018robust} & TDOF &
Estimated by maximum likelihood method &
A small number of degrees of freedom (df) value indicates a high frequency of extreme values (i.e heavy-tails), whereas a large df value signifies that the data distribution is approaching Gaussian one. \\ \hline
4& $2-\beta$, where $\beta$ is a shape factor of the generalized Gaussian distribution \cite{Nadarajah01092005,Iskander}&2-BETA&Estimated by maximum likelihood method&It controls the decay rate of the distribution's tails, where $\beta<2$ ($2-\beta>0$) characterize a "sharper" peak and heavier tails. \\\hline

\end{longtable}

\begin{longtable}{|c|m{3cm}|c|>{\centering\arraybackslash}m{4.5cm}|m{6cm}|}
\caption{Distribution-distance measures}
\renewcommand{\arraystretch}{0.8}
\label{tab4}\\
\hline
Id & Name & Abb & Formula / description & Interpretation \\
\hline
\endfirsthead
\hline
Id & Name & Abb & Formula / description & Interpretation \\
\hline
\endhead
\hline
\endfoot
\hline
\endlastfoot
1 & Kolmogorov--Smirnov statistic \cite{ks_test} & KS &
$\sup_i |\hat{F}(x_i) - F_0(x_i)|$ &
It quantifies the maximum distance between the empirical distribution of the observed data and a reference Gaussian CDF, where larger deviations directly indicate the presence of heavy-tailed behavior.\\ \hline
2 & Cram\'er--von Mises statistic \cite{cvm_test} & CvM &
$\sum\limits_{i=1}^{k} \left( \frac{2i-1}{2k} - F_0(x_{(i)}) \right)^2+\frac{1}{12k}$ 
&
It calculates the integrated squared difference between the empirical CDF of the signal and a reference Gaussian distribution, making it highly sensitive to the  deviations caused by large observations. \\ \hline
3 & Hellinger divergence \cite{grzesiek2021method} & HD &
$\frac{\sum\limits_{i=1}^{k}\left(\sqrt {\hat{f}(x_i)}-\sqrt{f_0(x_i)}\right)^2}{\sqrt2}$ &
It quantifies the similarity between two probability distributions (expressed by PDFs), specifically measuring how much the observed data's distribution "disjoins" from a reference Gaussian distribution due to extreme values. \\ \hline
4 & Jeffreys divergence \cite{said2017cluster} & JD &
$\sum\limits_{i=1}^{k}\left(\hat{f}(x_i)-f_0(x_i)\right)A(x_i)$, where $A(x_i)=\log\frac{\hat{f}(x_i)}{f_0(x_i)}$ &
JD (also known as the symmetrized Kullback-Leibler divergence) quantifies the statistical "distance" between the observed signal distribution and a reference Gaussian distribution in the means of PDFs. \\ \hline






\end{longtable}

\begin{longtable}{|c|m{3cm}|c|>{\centering\arraybackslash}m{4cm}|m{6.5cm}|}
\caption{Test statistics}
\renewcommand{\arraystretch}{0.8}
\label{tab5}\\
\hline
Id & Name & Abb & Formula / description & Interpretation \\
\hline
\endfirsthead
\hline
Id & Name & Abb & Formula / description & Interpretation \\
\hline
\endhead
\hline
\endfoot
\hline
\endlastfoot

1 & Jarque--Bera statistic \cite{JB_test} & JB &
$\frac{k}{6}\left(SKEW^2+\frac{(KURT-3)^2}{4}\right)$ &
It is used to determine whether a given signal has the skewness and kurtosis matching a Gaussian distribution. \\ \hline




2 & Shapiro--Wilk statistic \cite{yap2011comparisons} & SW &
$\frac{\left(\sum\limits_{i=1}^{k} a_i x_{(i)}\right)^2}{\sum\limits_{i=1}^{k} (x_i-\bar x)^2}$, where $a_i$ are specific weights &
Assesses departures from Gaussian distribution by detecting deviations in order statistics, especially skewness and kurtosis. \\ \hline


3 & Epps--Singleton statistic \cite{es_test} & ES &
$(\hat{\theta} - \theta_0)^\top \hat\Sigma^{-1} (\hat{\theta} - \theta_0)$, where $\hat\Sigma^{-1}$ is a inverse of appropriate covariance matrix & It compares empirical and theoretical (for Gaussian distribution) characteristic functions, making it sensitive to deviations from basic distribution. \\ \hline
4& Modified Greenwood statistic \cite{gmodified_greendwood}&MGS& $\frac{\sum\limits_{i=1}^{k} |x_i|^2}{\left(\sum\limits_{i=1}^{k} |x_i|\right)^2}$ &It quantifies the "clumpiness" or unevenness of energy distribution within a signal, specifically by looking at the squared values of the sample points.\\\hline
\end{longtable}

\section{Bootstrap for dependent data} \label{app:bootstrap_app}
This appendix describes the circular block bootstrap procedure used to generate resampled datasets while preserving the dependence structure of cyclic data.

Let $\mathbf{x} :=  x_1, \dots, x_{k}$  denote the original data sequence of length $k$. The goal is to construct $B$ bootstrap samples, each of length $m$, using blocks of consecutive observations of fixed size $L$. The circular block bootstrap operates by repeatedly sampling contiguous blocks from the original series, allowing wrap-around at the boundaries to maintain cyclic consistency. The procedure is defined as follows:

\begin{enumerate}
    \item Initialize a pseudo-random number generator with a fixed seed to ensure reproducibility.
    \item For each bootstrap replication $b = 1, 2, \dots, B$:
    \begin{enumerate}
        \item Initialize an empty sequence $\mathbf{x}^{*(b)}$.
        \item While $|\mathbf{x}^{*(b)}| < m$:
        \begin{enumerate}
            \item Draw a starting index $s \sim \mathcal{U}\{1, \dots, k\}$.
            \item Construct a block of length $L$:
            \begin{equation}
                \mathbf{z} = \{x_{(s + i) \bmod k} \mid i = 0, 1, \dots, L-1\}.
            \end{equation}
            \item Append $\mathbf{z}$ to $\mathbf{x}^{*(b)}$.
        \end{enumerate}
        \item Truncate $\mathbf{x}^{*(b)}$ to length $m$ if necessary.
    \end{enumerate}
    \item Store all bootstrap samples in a matrix $\mathbf{X}^* \in \mathbb{R}^{B \times k}$.
\end{enumerate}

The use of modular indexing ensures that blocks can wrap around the end of the sequence, which is particularly suitable for periodic or cyclic data. By sampling blocks instead of individual observations, the method preserves local correlation structures up to lag $L-1$. The choice of block size $L$ controls the bias-variance trade-off: larger blocks better preserve dependence but reduce variability across bootstrap samples. The resulting matrix $\mathbf{X}^*$ contains $B$ bootstrap realizations that can be used for statistical inference, such as estimating confidence intervals or variability measures under dependence.

Due to the fact that the algorithm operates on blocks rather than individual observations, it is advisable in practice to increase the length of the reference dataset from which samples are drawn. A longer input sequence allows for a greater number of distinct block combinations, thereby improving the diversity of the generated bootstrap samples. In particular, when the data length $n$ is small relative to the block size $L$, the number of unique blocks that can be formed is limited, which may lead to repetitive patterns in the resampled sequences. This can reduce the effectiveness of the bootstrap in approximating the sampling distribution. Therefore, extending the reference dataset - when possible - enhances the variability and representativeness of the bootstrap replicates.

In Section \ref{sec:impulsivity_assessment} the reference signals had a duration of 10s, while the test signals were 1s long. The test statistic was computed from the test signal with respect to the last 10s segment of the reference signal (ensuring equal data lengths). 
\newpage\clearpage
\section{Additional figures for metrics selection and additional results of $U^*$-based performance}
\label{app:figures}

\begin{figure}[h!]
    \centering
    \includegraphics[width=0.49\textwidth]
    {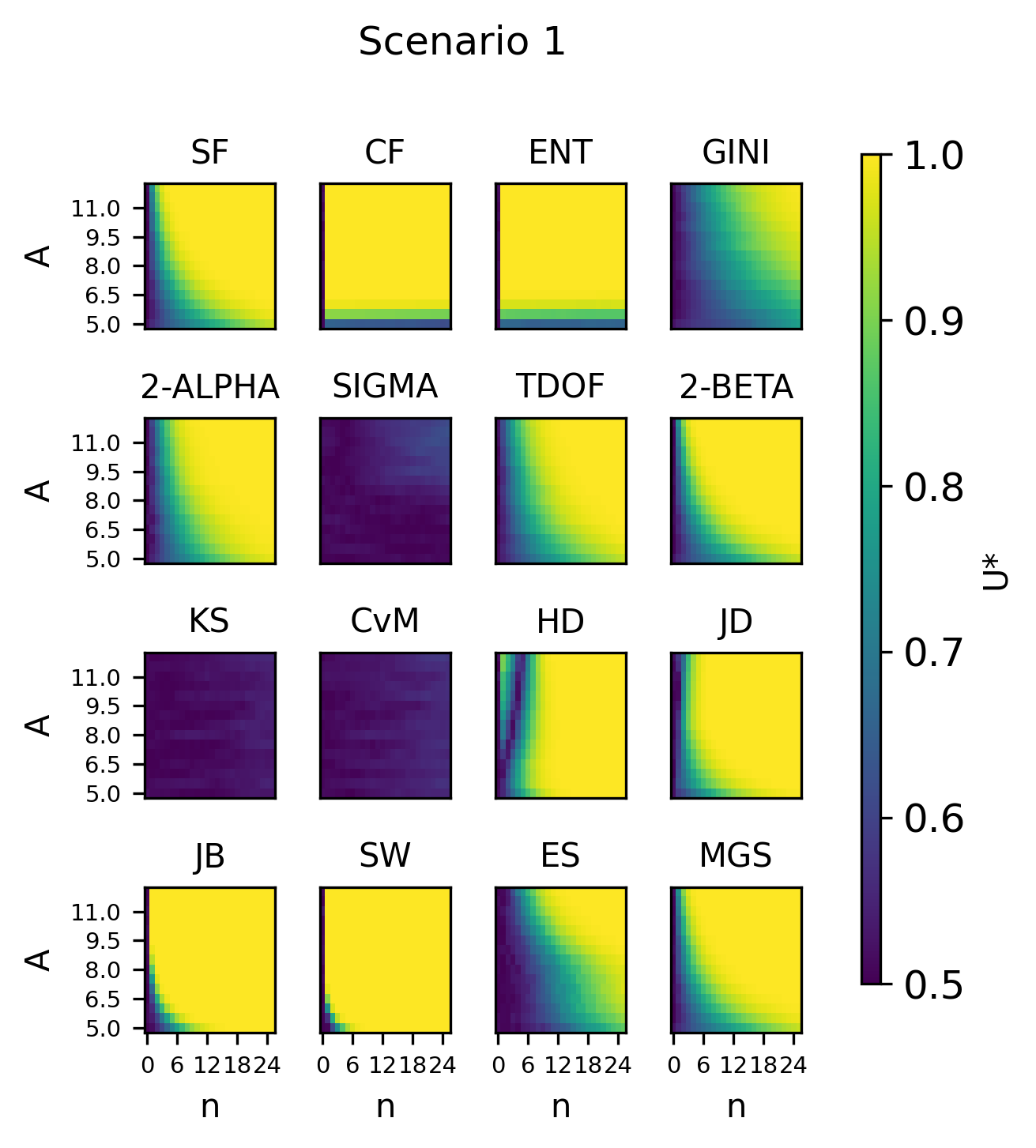}
    \includegraphics[width=0.49\textwidth]{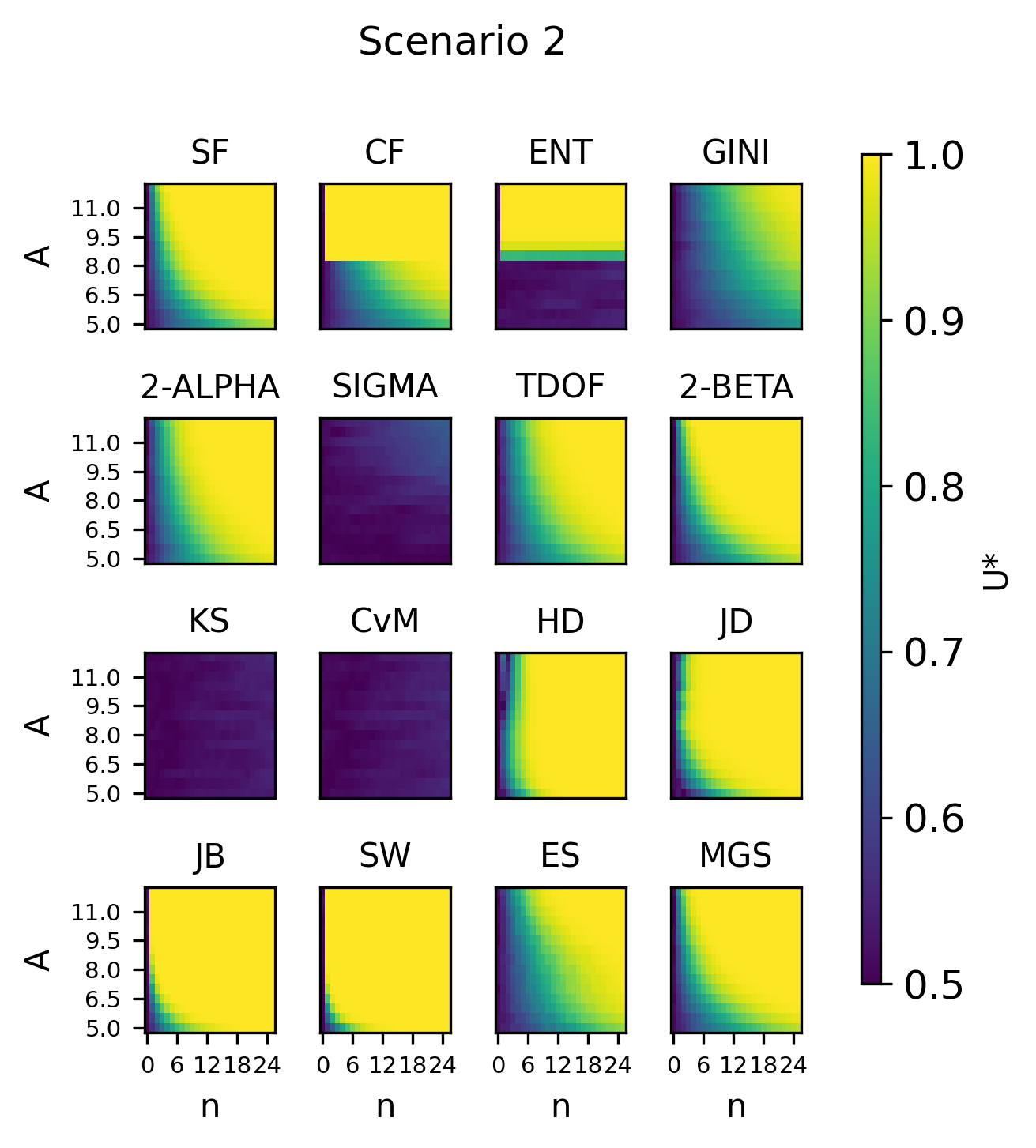}
    \includegraphics[width=0.49\textwidth]{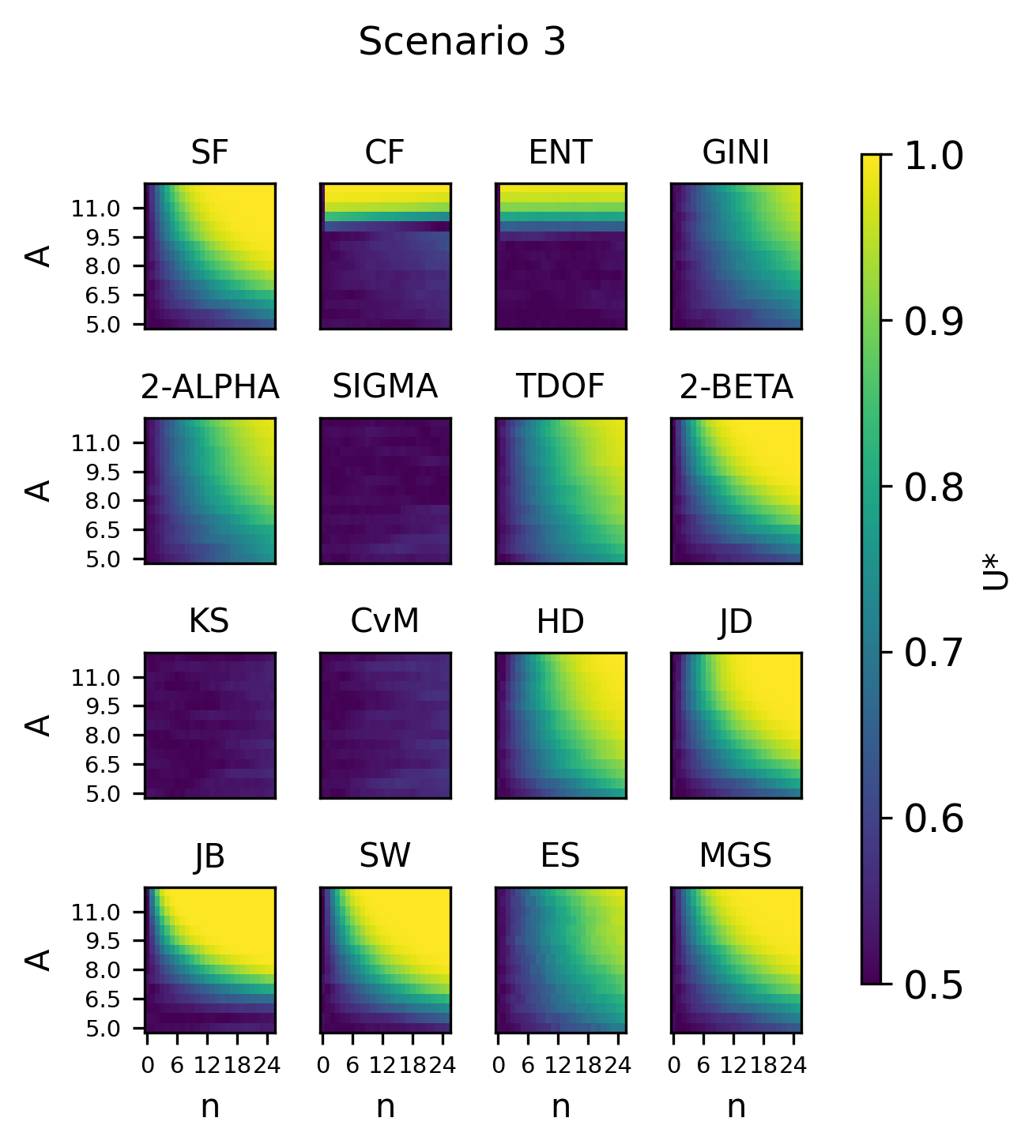}
    \caption{Comparison of $U^*$ heatmaps for all scenarios and for the impulsivity metrics presented in Tables \ref{tab2}-\ref{tab5}}
\label{fig:heatmap4x4_scenario1}
\end{figure}


\begin{table}[htbp]
\centering
\renewcommand{\arraystretch}{0.8}
\caption{$U^*$-based performance metrics for all scenarios.}
\label{tab:u_star_all}
\begin{tabular}{l c c c c c}
\toprule
Metric & Mean imp. & Mean hard & Mean easy & Coverage & Worst case \\
\midrule

\multicolumn{6}{c}{\textbf{Scenario 1}} \\
\midrule
SF & 0.930 & 0.884 & 0.971 & 0.818 & 0.540 \\
CF & \textbf{0.968} & \textbf{0.932} & \textbf{1.000} & \textbf{0.897} & 0.617 \\
ENT & 0.966 & 0.928 & \textbf{1.000} & \textbf{0.897} & \textbf{0.649} \\
GINI & 0.773 & 0.712 & 0.826 & 0.433 & 0.509 \\
\addlinespace
2-ALPHA & 0.908 & 0.872 & 0.939 & 0.772 & 0.518 \\
SIGMA & 0.531 & 0.510 & 0.550 & 0.000 & 0.500 \\
TDOF & 0.896 & 0.856 & 0.930 & 0.744 & 0.539 \\
2-BETA & \textbf{0.930} & \textbf{0.884} & \textbf{0.971} & \textbf{0.821} & \textbf{0.541} \\
\addlinespace
KS & 0.520 & 0.519 & 0.521 & 0.000 & 0.500 \\
CvM & 0.534 & 0.533 & 0.535 & 0.000 & 0.500 \\
HD & 0.919 & 0.923 & 0.915 & 0.785 & \textbf{0.514} \\
JD & \textbf{0.949} & \textbf{0.935} & \textbf{0.962} & \textbf{0.862} & 0.508 \\
\addlinespace
JB & 0.982 & 0.962 & \textbf{1.000} & 0.923 & 0.517 \\
SW & \textbf{0.994} & \textbf{0.986} & \textbf{1.000} & \textbf{0.949} & 0.507 \\
ES & 0.793 & 0.724 & 0.854 & 0.518 & 0.500 \\
MGS & 0.930 & 0.884 & 0.971 & 0.818 & \textbf{0.540} \\

\midrule
\multicolumn{6}{c}{\textbf{Scenario 2}} \\
\midrule
SF & \textbf{0.928} & \textbf{0.881} & 0.970 & \textbf{0.815} & \textbf{0.537} \\
CF & 0.906 & 0.798 & \textbf{1.000} & 0.751 & 0.530 \\
ENT & 0.766 & 0.527 & 0.975 & 0.513 & 0.501 \\
GINI & 0.770 & 0.711 & 0.822 & 0.436 & 0.518 \\
\addlinespace
2-ALPHA & 0.911 & 0.877 & 0.941 & 0.782 & \textbf{0.539} \\
SIGMA & 0.540 & 0.515 & 0.563 & 0.000 & 0.500 \\
TDOF & 0.890 & 0.850 & 0.925 & 0.736 & 0.543 \\
2-BETA & \textbf{0.928} & \textbf{0.880} & \textbf{0.970} & \textbf{0.808} & 0.537 \\
\addlinespace
KS & 0.523 & 0.522 & 0.524 & 0.000 & 0.500 \\
CvM & 0.526 & 0.525 & 0.527 & 0.000 & 0.500 \\
HD & 0.952 & \textbf{0.954} & 0.950 & 0.862 & \textbf{0.503} \\
JD & \textbf{0.957} & 0.934 & \textbf{0.977} & \textbf{0.885} & 0.502 \\
\addlinespace
JB & 0.985 & 0.969 & \textbf{1.000} & 0.933 & 0.568 \\
SW & \textbf{0.992} & \textbf{0.983} & \textbf{1.000} & \textbf{0.946} & \textbf{0.576} \\
ES & 0.861 & 0.812 & 0.904 & 0.664 & 0.533 \\
MGS & 0.928 & 0.881 & 0.970 & 0.815 & 0.537 \\

\midrule
\multicolumn{6}{c}{\textbf{Scenario 3}} \\
\midrule
SF & \textbf{0.801} & \textbf{0.691} & \textbf{0.896} & \textbf{0.508} & \textbf{0.500} \\
CF & 0.645 & 0.534 & 0.743 & 0.221 & 0.501 \\
ENT & 0.627 & 0.509 & 0.730 & 0.192 & 0.500 \\
GINI & 0.705 & 0.642 & 0.759 & 0.256 & 0.500 \\
\addlinespace
2-ALPHA & 0.744 & 0.684 & 0.796 & 0.344 & \textbf{0.511} \\
SIGMA & 0.516 & 0.524 & 0.509 & 0.000 & 0.500 \\
TDOF & 0.769 & 0.726 & 0.807 & 0.433 & 0.500 \\
2-BETA & \textbf{0.800} & \textbf{0.690} & \textbf{0.896} & \textbf{0.505} & 0.500 \\
\addlinespace
KS & 0.520 & 0.520 & 0.519 & 0.000 & 0.500 \\
CvM & 0.535 & 0.537 & 0.534 & 0.000 & 0.501 \\
HD & 0.793 & \textbf{0.739} & 0.841 & 0.500 & \textbf{0.502} \\
JD & \textbf{0.815} & 0.727 & \textbf{0.892} & \textbf{0.541} & 0.500 \\
\addlinespace
JB & 0.798 & 0.633 & \textbf{0.942} & 0.521 & 0.500 \\
SW & \textbf{0.812} & 0.686 & 0.923 & \textbf{0.544} & \textbf{0.502} \\
ES & 0.732 & 0.679 & 0.779 & 0.328 & 0.501 \\
MGS & 0.801 & \textbf{0.691} & 0.896 & 0.508 & 0.500 \\

\bottomrule
\end{tabular}
\end{table}

\clearpage
\newpage
\section{Additional figures for impulsiveness assessments} \label{app:figures_imp_assessment}

\begin{figure}[h!]
    \centering
    \includegraphics[width=\textwidth]{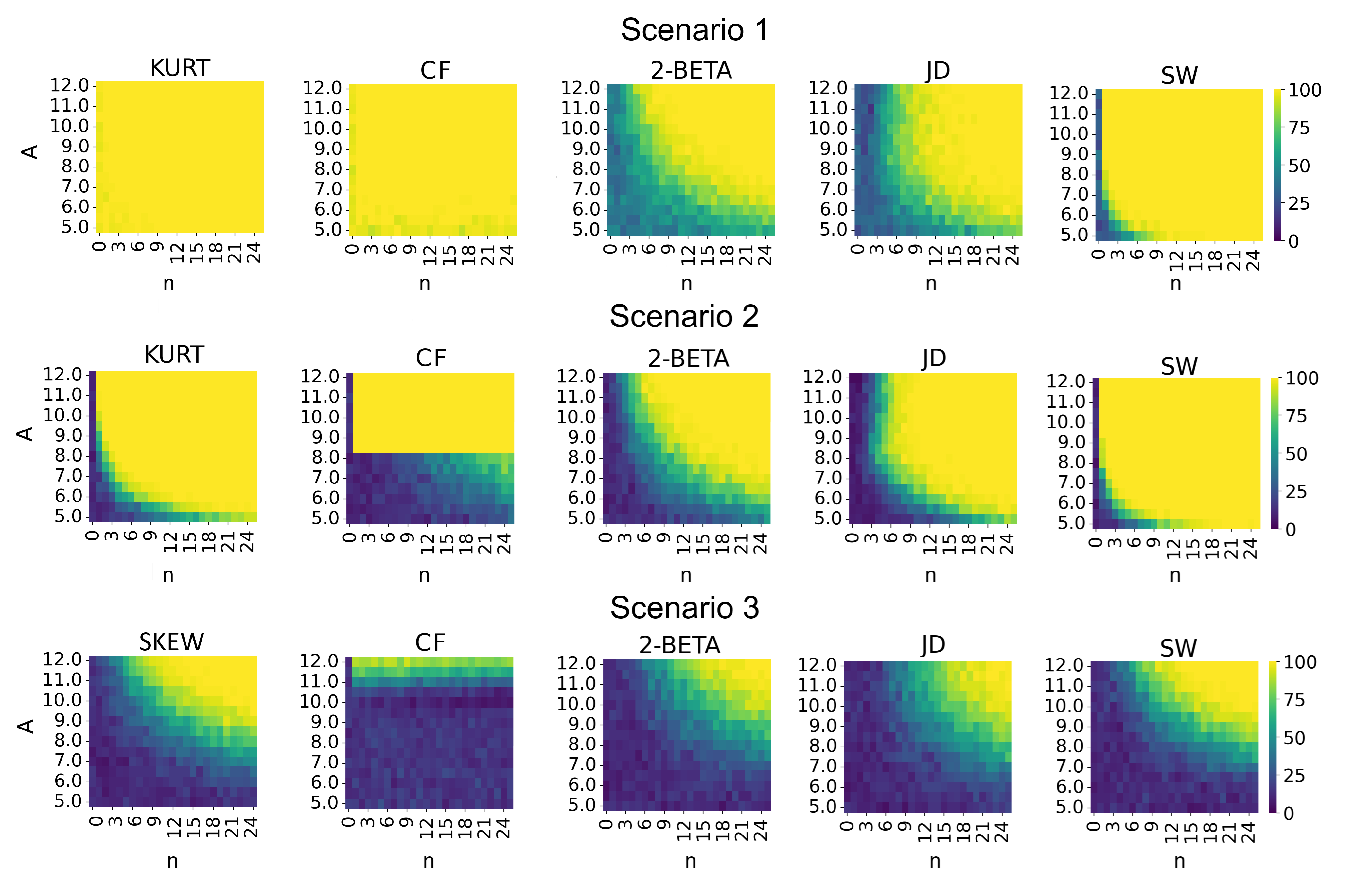}
    \caption{Percentage of Monte Carlo trials in which the bootstrapping-based methodology, according to the selected criterion, indicates a significant change in impulsiveness between the test signal and the reference signal.}
    \label{fig:s1_ap}
\end{figure}

\end{document}